\documentclass[twocolumn, showpacs,amsfonts,amsmath,floatfix, superscriptaddress, prx]{revtex4-2}
\usepackage{graphicx}
\usepackage{import}
\usepackage{color,verbatim,bbold, float}
\usepackage{amssymb,amsbsy,mathrsfs}
\usepackage{times}
\usepackage{color}
\usepackage{setspace}
\usepackage{bm}% bold math.
\usepackage{braket}
\usepackage{commath}
\usepackage[dvipsnames]{xcolor}
\usepackage{csquotes}
\usepackage{enumitem}
\usepackage{dsfont}
\usepackage{appendix}
\usepackage{xcolor}
\usepackage{comment}
\usepackage{mathtools}
\usepackage{calc}
\usepackage[colorlinks,allcolors=blue]{hyperref}
\usepackage{physics}

\newcommand{\commad}[1]{{\textcolor{red} {\it{[Note (Adithi): #1]}}}} %to insert my comments

\def\l{\left} 
\def\r{\right}
\def\nn{\nonumber}
\def\d{\dagger}

\begin{document}
%\title{Characterisation of non-Markovian effects induced by random telegraph noise in bosonic channels}
\title{Performance of rotation-symmetric bosonic codes in the presence of non-Markovian effects induced by random telegraph noise}

\author{Adithi Udupa}
\affiliation{Department of Microtechnology and Nanoscience, Chalmers University of Technology, Göteborg SE-412 96, Sweden}
\author{Timo Hillmann}
\affiliation{Department of Microtechnology and Nanoscience, Chalmers University of Technology, Göteborg SE-412 96, Sweden}
\author{Rabsan Galib Ahmed}
\affiliation{Department of Physical Sciences, Indian Institute of Science Education and Research Mohali, Punjab 140306, India}
\author{Andrea Smirne}
\affiliation{Department of Physics, University of Milan, Milan 20122, Italy}
\affiliation{Istituto Nazionale di Fisica Nucleare, Sezione di Milano, Via Celoria 16, I-20133 Milan, Italy}
\author{Giulia Ferrini}
\affiliation{Department of Microtechnology and Nanoscience, Chalmers University of Technology, Göteborg SE-412 96, Sweden}

\begin{abstract}
Decoherence in quantum devices, such as qubits and resonators, is often caused by bistable fluctuators modeled as random telegraph noise (RTN), leading to significant dephasing. 
We analyze the impact of individual and multiple fluctuators on a bosonic mode in continuous variable systems, identifying non-Markovian behavior governed by two timescales: the switching rate ($\xi$) and the coupling strength ($\nu$) of the fluctuator. Using the Breuer-Laine-Piilo (BLP) trace-distance measure, we characterize non-Markovianity for both Gaussian and non-Gaussian states, revealing that for rotation-symmetric bosonic (RSB) codes, known for their error correction advantages, the measure grows linearly with code symmetry and can become unbounded. We evaluate the performance of these RSB codes under simultaneous loss and RTN dephasing.
For a teleportation-based Knill error-correction circuit, the codes perform robustly in the Markovian limit. 
In the non-Markovian regime, the performance depends non-trivially on the time at which the error correction is performed.
The average gate fidelity of the error-corrected state in this case exhibits oscillations as a function of time due to the oscillatory nature of the dephasing function of the RTN noise; however, for most of the parameter ranges, the values stay beyond the break-even point. Extending to multiple fluctuators that produce $1/f$ noise, we observe that non-Markovianity decays with increasing fluctuator count, while the performance of RSB codes remains effective with increasing number of fluctuators.

\end{abstract}

\maketitle

\section{Introduction}

Bosonic codes are promising for hardware-efficient quantum error correction (QEC) protocols and hold thereby significant potential for achieving fault-tolerant quantum computation~\cite{Noh-PRA-2020, Tzitrin2021, Grimsmo2021, gkpxanadu, gouzien2023performance, guillaud2019repetition, Darmawan_2021, lemonde_hardware-efficient_2024, aghaee2025scaling, walshe_linear-optical_2025, hann_hybrid_2024} with reduced overheads. 
Implementations of bosonic codes for QEC have been achieved with superconducting microwave circuits~\cite{sivak2023real, ni2023beating, putterman_hardware-efficient_2025, marquet_autoparametric_2024, reglade_quantum_2024,rousseau_enhancing_2025, brock_quantum_2024},
% ~\commg{Timo: are there experimental references from Alice and Bob?}, 
trapped ions~\cite{fluhmann2019encoding, de2022error, matsos_robust_2024, matsos_universal_2024} and on photonic platforms~\cite{konno2024logical, aghaee2025scaling}.
% ~\commg{Timo: More references?}

Along with photon or phonon losses, one of the main sources of noise in bosonic systems, and in some cases the dominating one~\cite{fluhmann2019encoding}, is (photon number) dephasing. %Dephasing even dominates over relaxation in trapped ion setups, making it the primary channel of decoherence. 
For dephasing type of noise, bosonic codes with a discrete rotational symmetry in phase space are particularly suitable and yield good performance in terms of the capability of restoring the intended quantum information~\cite{qec3, timo, channel}, due to phase space structure of the codewords, i.e., the bosonic states which encode the logical qubit states $\ket{0}$ and $\ket{1}$~\cite{arne}. %, which are characterized by a well-defined position in phase space.  

So far, the performance of rotationally symmetric bosonic codes (RSB) under dephasing noise has been studied only for the simplest case of Gaussian, Markovian dephasing~\cite{timo, channel,ouyang2021trade}. 
However, in several experimental platforms, other types of noise arise.
For instance, microwave cavities coupled to superconducting qubits are affected by decoherence effects stemming from the coupling with one or more random fluctuators present on the sample substrate~\cite{exptls1, exptls2, exptls3, exptls4, exptls5, exptls3d}.
These fluctuators are often modeled in terms of two-level systems (TLSs)~\cite{tls1, tls2, tls3, tls4} switching between two metastable states at random intervals and yielding a random telegraph noise (RTN)~\cite{decohrtn1, decohrtn2}.
The dynamics of a state coupled to such a TLS with an average switching rate $\xi$ and coupling strength $\nu$ undergoes a transition from a Markovian to a non-Markovian evolution depending on the ratio of the two timescales, $r=\xi/\nu$.
RTN is also often associated with a colored noise with characteristic $1/f$ power spectrum ~\cite{decohrtn3, decohrtn4}. This is modeled by either using a distribution of switching rates for a single RTN fluctuator, or by coupling multiple independent fluctuators to the system with different switching rates.
 %When the quantum information is encoded in motional modes (i.e., harmonic oscillator states) of the ion, slow, correlated noise in electric fields or trap parameters can cause non-Markovian dephasing in the oscillator's phase space~\cite{trappedion2, trappedion3}. 

The performance of bosonic codes under these types of noise has not yet been studied; indeed, while error correction under non-Markovian noise has been studied for qubits~\cite{biswas2024, kam2024}, it remains largely unexplored for bosonic systems. 
Furthermore, so far, the study of non-Markovianity in bosonic systems from an open quantum system perspective, e.g., when it comes to quantifying non-Markovianity with suitable measures, has been limited to the case of Gaussian states and Gaussian channels~\cite{fidelity, Gaussian_channels}.

In this work, we address both gaps.
On the one hand, we study the performance of RSB under RTN and colored noise with a $1/f$ spectrum for the case of a realistic state recovery procedure. The state recovery procedure that we consider is based on the Knill teleportation scheme, and its constituent circuit elements are readily available in photonics and superconducting labs~\cite{arne, timo}. In the Markovian limit of the RTN, the RSB codes are found to perform well, agreeing with the previous results in Refs.~\cite{arne, timo}, including the case where dephasing is combined with a loss channel. In the non-Markovian limit, we see that the performance depends on the time at which error correction happens, with certain timescales giving a significant improvement in fidelities compared to the break-even potential, the threshold where the lifetime of the error-corrected qubit is as good as that of an unencoded qubit. We show that the span of these desirable time intervals can be further increased by tuning the parameters of the RSB codes. For a given time, however, the performance degrades close to the transition between the two regimes. 

On the other hand, we also characterize the non-Markovianity arising in a bosonic mode coupled to non-Gaussian classical environments corresponding to the two types of noise described above by evaluating the trace distance between two initial states for the case of rotationally symmetric bosonic codewords, which are a subset of non-Gaussian states. We show that the resulting non-Markovianity measure exceeds the one obtained for the case of Gaussian states and that it increases linearly with the order of RSB code symmetry, suggesting that it is unbounded.

Our results have therefore profound implications both for identifying codes and parameter regimes yielding the most effective quantum error correction in the presence of non-Markovian dephasing and for understanding and characterizing such types of noise from an open quantum system perspective.

The structure of the rest of the paper is as follows: In Sec.~\ref{sec:summary}, we summarise the main results of the paper.
The terminology and basic concepts for both QEC and open quantum systems used therein will be introduced in more detail in the following sections. Sec.~\ref{sec:rtn_intro} provides a detailed analysis of the RTN and $1/f$ noise models in bosonic modes, and presents simulations and analytical calculations for the dynamics of the state of a bosonic mode under such noises, in both Markovian and non-Markovian regimes. In Sec.~\ref{sec:quantifying_N_blp}, we assess the non-Markovianity measure for Gaussian states and rotation-symmetric bosonic states, respectively, in the presence of RTN. 
Finally, in Sec.~\ref{sec:performance}, we analyze the performance of rotation-symmetric bosonic codes such as cat and binomial codes under a Knill error correction circuit, with the input states subjected to loss and RTN dephasing or $1/f$ noise.
All the numerical calculations have been performed using the QuTip package~\cite{qutip1, qutip2}. Appendices \ref{app:dephasing_function}-\ref{app:fidelity_derivation} address various technical details connected to the results above. %In Appendix~\ref{app:microscopic_model}, we give a brief description of the microscopic model of the Hamiltonian and derivation 
In Appendix~\ref{app:dephasing_function} and \ref{app:multiple_fluctuator_derivation}, we derive the dephasing function for RTN and $1/f$ noise and the resulting evolution of the density matrix, respectively. We further show the analytic derivations of the evolution of the density matrix in the two limits of the RTN switching rates in Appendix~\ref{app:two_limits}. 
In Appendix \ref{app:coherent_state_maximisation}, we detail the optimization procedure to obtain a lower bound on the non-Markovian measure based on trace distance for the case of Gaussian states, and in Appendix~\ref{app:nblp_non_gaussian}, we show analytically why this measure is unbounded for non-Gaussian states. 
In Appendix~\ref{app:wigner_neg}  we show that another non-Markovianity measure, based on Wigner negativity for non-Gaussian states, qualitatively matches the behavior of the trace-distance-based measure. 
The derivation of the recovery map for teleportation-based Knill-type error correction is presented in detail in Appendix~\ref{app:circuit_derivation}. To the best of our knowledge, this derivation has not been previously documented. 
In Appendix~\ref{app:fidelity_derivation}, we derive an expression for the average gate fidelity for a purely dephasing channel, aiding in the understanding of the numerical results obtained from the error correction protocol. Finally, in Appendix~\ref{app:1/f}, we present additional results of the characterization of non-Markvonianity and performance of binomial codes in the presence of $1/f$ noise. 
%in different regimes identified by the interplay of two timescales: the average switching rate of RTN $\xi$ and the coupling strength $\nu$ to the bosonic mode.
%. We analyze the revivals in trace distance in the highly non-Markovian regime $r= \xi/\nu \ll 1$. 
%We show that for the subset of Gaussian states, squeezing and thermal fluctuations do not increase the non-Markovianity measure, providing a bound on this measure for a given ratio of $r$ for Gaussian states. We also examine a model with multiple RTN fluctuators coupled to our bosonic mode, which mimics a $1/f$ power spectrum, and we show that non-Markovianity measures decrease exponentially with an increasing number of fluctuators. Additionally, we explore non-Markovian effects in specific classes of non-Gaussian bosonic codes, such as cat and binomial codes—referred to as rotation symmetric bosonic codes~\cite{qec3}, characterized by discrete rotation symmetry in phase space. Our findings reveal that the non-Markovian measure for these codes increases linearly with the order of symmetry, suggesting that it is unbounded. 

\section{Summary of main results}
\label{sec:summary}
%We present a comprehensive study of dephasing induced by random telegraph noise (RTN) from one or more two-level systems (TLS) coupled to a bosonic code. The analysis spans different regimes defined by the characteristic parameters of RTN. A single RTN source is characterized by its average switching rate $\xi$ and coupling strength $\nu$ to the bosonic mode. Varying the ratio $r=\xi/\nu$ reveals a transition in system dynamics from non-Markovian to Markovian behavior at a given time. To quantify non-Markovianity, we adopt a widely used measure based on the trace distance evolution between quantum states~\cite{tracedistance}.

%In QEC studies, the Gaussian dephasing channel which is Markovian in nature, along with a photon loss channel~\cite{channel} are often employed in simulating noisy environments so far. Prior works in~\cite{arne} and~\cite{timo} have examined the performance of symmetric bosonic codes~\cite{qec3} under Gaussian dephasing and photon loss. However, a more realistic modeling of dephasing due to TLS—common in experimental platforms—has been explored only in for the qubit QEC protocols~\cite{biswas2024, kam2024}. Motivated by this, 

\begin{figure*}
\centering
\includegraphics[width=\linewidth]{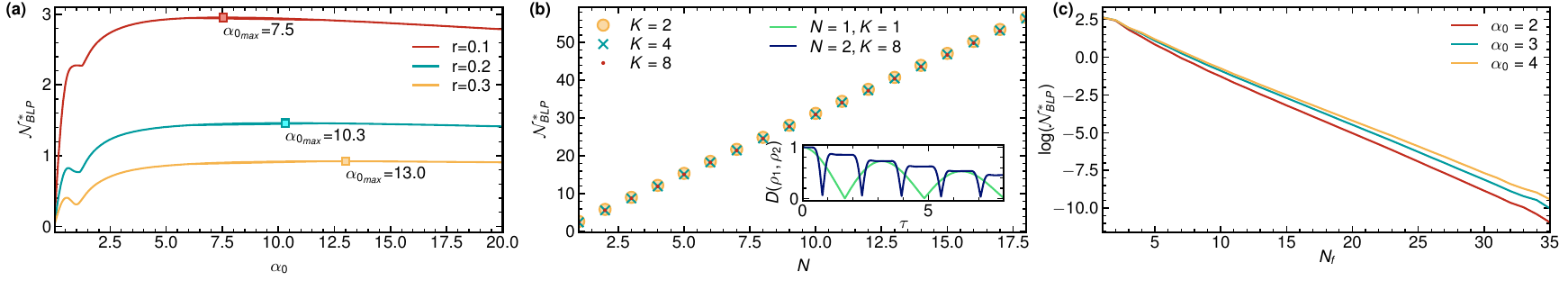}
\caption{ (a) For Gaussian states, we show that for a given pair of coherent states with $\alpha_0$ and $-\alpha_0$, the value of $\alpha_0$ that maximizes $\mathcal{N}_{\text{BLP}}^{*}$ depends on $r$. (b) Non-Markovianity measure for binomial codewords (non-Gaussian states) as a function of the order of rotation symmetry $N$. The measure linearly increases with $N$, due to the linear increase of frequency of oscillations in the trace distance between the codewords (shown in the inset) (c) The behavior of \(\mathcal{N}_{\mathrm{BLP}}^{*}\) as a function of the number of fluctuators \(N_f\) in the simulation of $1/f$ noise. For all considered values of \(\alpha = 2, 3, 4\) for coherent states, \(\mathcal{N}_{\mathrm{BLP}}^{*}\) decreases exponentially with \(N_f\), indicating a transition toward a Markovian regime. }
\label{fig:main_results2}
\end{figure*}

We first take an open quantum system perspective and quantify the non-Markovian behavior induced by RTN using a trace distance-based measure, due to Breuer, Leine, and Piilo (BLP), $\mathcal{N}_{\text{BLP}}$. We derive a lower bound on such a non-Markovianity measure for a broad class of Gaussian states (Fig.~\ref{fig:main_results2}(a)) and observe that there exist non-Gaussian CV states, such as the codewords of RSB codes, for which this measure is unbounded (Fig.~\ref{fig:main_results2}(b)). We also analyze the dependence of non-Markovianity on code parameters such as rotational symmetry and average photon number of the binomial and cat states, respectively. For $1/f$ noise modeled via multiple RTN sources, the system undergoes a crossover to Markovian dynamics when the number of coupled fluctuators exceeds approximately ten (for initial states taken as coherent states, Fig.~\ref{fig:main_results2}(c)).

In more detail, for Gaussian states, numerical simulations show that squeezing and thermal noise do not affect non-Markovianity. For fixed \( r \), the measure is maximized at an optimal coherent state amplitude \( \alpha_0 \), as shown in Fig.~\ref{fig:main_results2}(a). For codewords of rotation-symmetric bosonic, we find that the non-Markovianity measure increases approximately linearly with the rotational symmetry order \( N \), driven by a corresponding increase in the oscillation frequency of the trace distance. Figure~\ref{fig:main_results2}(b) shows this trend in the non-Markovian regime (\( r < 1 \)) for a binomial code with parameters \( N \) and \( K \). As illustrated in the inset, increasing \( N \) enhances oscillation frequency, thereby boosting the measure, whereas increasing \( K \) primarily broadens the oscillation peaks with negligible effect on non-Markovianity. Figure~\ref{fig:main_results2}(c) shows analogous behavior for coherent state pairs (\( \alpha_0 = 3 \), relative phase \( \pi \)) under \( 1/f \) noise, modeled via \( N_f \) fluctuators (see Sec.~\ref{sec:evolution-1overf}). As \( N_f \) exceeds 10, trace distance oscillations decay, signaling a crossover to Markovian dynamics.

 \begin{figure*}
    \centering
    \includegraphics[width=\linewidth]{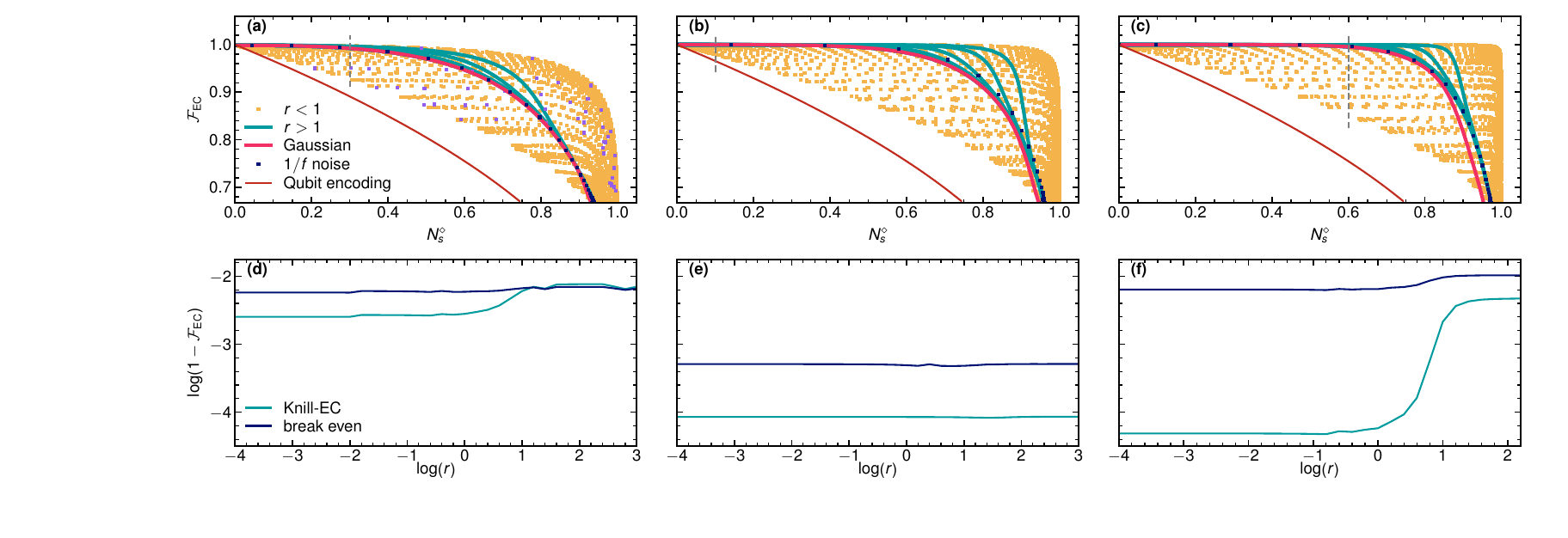}
\caption{
Upper panel: Average gate fidelity \(\mathcal{F}_{\text{EC}}\) of the error-corrected output state from the Knill-EC circuit with canonical phase measurements (details in Sec.~\ref{sec:performance}) as a function of the noise strength in the input state, induced by dephasing from RTN or \(1/f\) noise, for binomial encoding with parameters: (a) \(N = 2, K = 5\); (b) \(N = 3, K = 8\); and (c) for cat encoding with \(N=2, \alpha^2 = 12\).
The noise strength $N_s^{\diamond}$ is quantified using the diamond norm of the noise map projected onto the code space at different time points.
The orange squares represent the non-Markovian regime with ratio \(r < 1\). The fidelity  \(\mathcal{F}_{\text{EC}}\) for this case exhibits non-monotonic behavior with respect to the noise strength due to the oscillatory nature of the dephasing function, that is,  we plot in purple squares for one such value of $r=0.1$ in the non-Markovian regime in (a). In contrast, for \(r > 1\), a monotonic decrease in \(\mathcal{F}_{\mathrm{EC}}\) is observed as noise strength increases. 
The \(1/f\) noise, modeled using ten fluctuators with switching rates taken from a distribution, exhibits behavior similar to that of the Gaussian case. \newline Lower panel: Infidelity of the recovered state as a function of $r$ along a fixed line (shown in dotted grey in the plots of the upper panel) is given below for (d) $N_s^{\diamond}=0.3$  (e) $N_s^{\diamond}=0.1$ and (f) $N_s^{\diamond}=0.6$.
These results are compared with the break-even threshold, and we observe that the codes perform better than break-even for most values of $r$. The performance in (d) in the Markovian limit can be further improved by increasing the value of the parameter $K$. 
}
    \label{fig:noise_strength}
\end{figure*}

We then assess the performance of rotation-symmetric bosonic codes such as binomial and cat codes in counteracting the effect of RTN under Knill-type error correction (EC) circuit~\cite{arne}. This teleportation-based circuit uses a combination of controlled-rotation (CROT) gates and phase measurements to convert physical errors into correctable logical errors. As a metric of QEC performance, we compute the average gate fidelity of the final recovered state \cite{av_gate_fid}. We extend this analysis to the case of $1/f$ noise, prevalent in most quantum devices. 

In the upper panel of Fig.~\ref{fig:noise_strength}, we plot the average gate fidelity $\mathcal{F}_{EC}$ of the recovered state subjected to Knill-EC circuit as a function of noise strength for various RTN parameters. The noise strength is characterized using the diamond norm~\cite{Iyer} of the noisy channel map after being projected onto the codespace (see Sec.~\ref{sec:circuit_details} for definitions and circuit details). We characterize the performance of binomial codes with different values of $N$ and $K$, representing the order of rotational symmetry and truncation limit, respectively (Fig.~\ref{fig:noise_strength}(a) and (b)) and cat codes for a given $N$ and $\alpha$, the parameter of the primitive coherent state. We show that in the non-Markovian regime ($r \ll 1$), the QEC performance strongly depends on periodically occurring time intervals. This regime displays non-monotonic fidelity behavior due to characteristic revivals coming from RTN dephasing. In contrast, in the Markovian regime ($r \gg 1$) and under $1/f$ noise for a large number of fluctuators, the fidelity exhibits a smooth and monotonic response to increasing noise strength. For comparison, we also consider the trivial encoding in a Fock qubit ($N =1, K = 1$), and find that encoding with higher $N$ consistently yields better performance in all the regimes. For each plot in the upper panel of Fig.~\ref{fig:noise_strength}, we provide a corresponding lower-panel plot of infidelity, defined as $1-\mathcal{F}_{EC}$, as a function of the ratio $r$ at a fixed noise strength, corresponding to the vertical lines in the upper panel. These results are further compared to the break-even threshold, beyond which bosonic encoding offers no advantage. We see that the infidelity of the error correction circuit depends on the parameters of the codewords but remains below break-even for most of the ranges of $r$, indicating enhanced QEC performance. For the case of $N=2$ in Fig.~\ref{fig:noise_strength}(a), although the performance becomes comparable to the break-even case for higher values of $r$, we do notice that it can be significantly improved by tuning the value of the parameter $K$- in this particular case by going from $K=5$ to $K=12$. Further details of the QEC performance of RSB codes under RTN are provided in Sec.~\ref{sec:numerical_results}.

Table \ref{tab:abbrevs} provides a summary of the abbreviations used in this work. We now proceed further to an extensive characterization of RTN and $1/f$ noises in bosonic systems. 
\begin{table}[H]
\setlength{\tabcolsep}{1pt}
\renewcommand{\arraystretch}{1.3}
\begin{ruledtabular}
 \begin{tabular}{c c}
        \textbf{Abbreviation} & \textbf{Definition} \\
        CV & Continuous Variable \\
        QEC & Quantum Error Correction \\
        RTN & Random Telegraph Noise\\
        BLP & Breuer, Laine and Piilo \\
        RSB & Rotation Symmetric Bosonic \\
        TLS & Two-Level System\\
    \end{tabular}
    \end{ruledtabular}
    \caption{List of abbreviations.}
    \label{tab:abbrevs}
\end{table}

\section{State evolution under Random Telegraph Noise and $1/f$ noise}
\label{sec:rtn_intro}
In the realization of solid-state qubits and superconducting microwave resonators, the primary intrinsic noise arises from fluctuations due to 
TLSs. These TLSs may result from tunneling atoms, dangling electronic bonds, impurity atoms, trapped charge states, and similar sources~\cite{tls1, tls2, tls3, tls4}. This noise can be attributed to a single strongly coupled fluctuator or a set of a few fluctuators with a distribution of hopping rates, generating $1/f$ noise. Such environments are often modeled by one or more fluctuators interacting with a bosonic bath, typically consisting of phonons or electron-hole pairs. The interaction between the fluctuators and the environment manifests as time-dependent random fields that affect the system, behaving as stochastic noise that can be treated classically~\cite{tls4}. 
In this section, we first investigate a bosonic mode coupled with a single fluctuator. We then analyze the non-Markovian signatures from the random telegraph noise for Gaussian and non-Gaussian states. We finally extend the study to a more generalized case of $1/f$ noise arising from either a single fluctuator with varying switching rates or from multiple fluctuators coupled to the bosonic mode.

\subsection{Hamiltonian and state evolution for RTN noise \label{sec:evolution-RTN}}
We first consider a bosonic mode coupled with strength $\nu$ to a single fluctuator that has an average switching rate $\xi$, resulting in random telegraph-like noise. As we will show, the coupling strength $\nu$ and the switching rate 
$\xi$ are the two critical timescales that determine the system's non-Markovianity. 
Let $c(t)$ represent the function that characterizes the fluctuator's state at time $t$, with value $\pm 1$ following the statistical properties of the RTN. This noise is further characterized by an exponentially decaying autocorrelation function and has a Lorentzian power spectrum. The Hamiltonian for the bosonic mode coupled to the time-dependent fluctuating field is given by \cite{tls1, tls4, quantumrtn, microscopic_model}
\begin{equation}
    \hat{H}_{\text{RTN}}= \epsilon \hat{a}^{\dagger}\hat{a} + \nu c(t)  \hat{a}^{\dagger}\hat{a},
    \label{eq:hamiltonian_bosonic}
\end{equation}
where $\epsilon$ is the energy level of the bosonic mode, and $\hat{a}$, $\hat{a}^{\dagger}$ are the annihilation and creation operators for the bosonic mode. The coupling strength of the bosonic mode to the field is given by $\nu$. We define the quantity $r$ to be the ratio of the two timescales $\xi$ and $\nu$, that is $r= \xi/\nu$. This interaction describes a non-dissipative dephasing channel, suitable for modeling environments where dephasing is the dominant decoherence mechanism. The effects of this type of noise have been well studied so far on qubits~\cite{qubit1, qubit2, qubit3, qubit4, bcm}. For a qubit coupled to a fluctuator, it has been shown that in the regime $r >1$, the dynamics are Markovian, while for  $r<1$, the system exhibits non-Markovianity.
Going back to our bosonic-mode system and
further introducing two dimensionless quantities $\tau = \nu t$ and $r=\xi/\nu$, we can express the evolution operator in the interaction picture as
\begin{equation}
    \hat{U}(t) = e^{i \phi_{\text{RTN}}(\tau) \hat{a}^\dagger \hat{a}},
\end{equation}
where the RTN phase $\phi_{\text{RTN}}(\tau)$ is defined as 
\begin{equation}
    \phi_{\text{RTN}}(\tau) = \int_0^\tau c(\tau^\prime) d\tau^\prime.
    \label{eq:rtnphase}
\end{equation}
The evolved density matrix is obtained by averaging the evolved density operator over the RTN phase Eq.~\eqref{eq:rtnphase},
\begin{equation}
    \hat{\rho}(\tau) = \langle \hat{U}(\tau) \hat{\rho}_0 \hat{U}^{\dagger}(\tau) \rangle_{\phi_{\text{RTN}}(\tau)}.
\end{equation}
Next, we consider the initial state $\hat{\rho}_0$ of the bosonic mode expressed in the Fock basis
\begin{equation}
    \hat{\rho}_0 = \sum_{m,n=0}^{\infty} \rho_{m,n} \ketbra{m}{n} ,
\end{equation}
where $\rho_{m,n}$ is the $(m,n)$-th matrix element of the density matrix. 
The evolved density matrix can thus be written as
\begin{equation}
    \hat{\rho}_{\text{RTN}}(\tau) = \sum_{m,n=0}^{\infty} \langle e^{i\phi_{\text{RTN}}(\tau) (m-n)} \rangle_{\phi_{\text{RTN}}(\tau)} \rho_{m,n} \ketbra{m}{n}.
    \label{eq:rho_evolved}
\end{equation}
The averaging $\langle .. \rangle_{\phi_{\text{RTN}}(t)}$ is performed over different instances of the random telegraph stochastic noise. 
The quantity $\langle e^{i\phi_{\text{RTN}}(\tau) (m-n)} \rangle_{\phi_{\text{RTN}}(t)}$ is called the dephasing function $G(r, \tau)$, 
and an analytical expression for it is derived in Appendix~\ref{app:dephasing_function} and reads
\begin{equation}
    \langle e^{i\phi_{\text{RTN}}(\tau) (m-n)} \rangle_{\phi_{\text{RTN}}(\tau)} = e^{-r \tau} (\cosh\Omega \tau + \frac{r}{\Omega} \sinh\Omega \tau),
\label{eq:dephasingfunction}
\end{equation}
where $\Omega = \sqrt{r^2 - (m-n)^2}$. This expression shows that, in addition to the ratio $r$, the dephasing function depends on the support of the state in the Fock basis, leading to either oscillatory or decaying behavior with time, depending on the given state. 

\subsection{Hamiltonian and state evolution for $1/f$ noise}
\label{sec:evolution-1overf}
To consider the impact of the interaction with more RTN fluctuators, we need to specify their spectral density. In numerous physical systems, the noise spectral density exhibits a $1/f$ dependence across a wide frequency range. This characteristic low-frequency behavior is observed in diverse systems, including bulk semiconductors, normal and superconducting metals, strongly disordered conductors, and devices fabricated from these materials~\cite{1/f1,1/f2,1/f3}. The simplest and most widely adopted model for describing $1/f$ noise involves a distribution of switching rates of single RTN fluctuators. For an individual RTN fluctuator, the power spectral density is Lorentzian and is expressed as~\cite{decohrtn3}:
 \begin{equation}
     S_{\xi}(\omega) = \frac{1}{\pi} \frac{2 \nu^2 \xi}{\omega^2 + (2\xi)^2}.
 \end{equation}

 The $1/f$ noise behavior can arise from either a single RTN fluctuator with a switching rate drawn from a statistical distribution $P(\xi)$ or a collection of RTN fluctuators with fixed but distinct switching rates, also distributed according to $P(\xi)$. In both cases, the combined spectral density over a frequency range from $\xi_{\rm min}$ to $\xi_{\rm max}$ is given by 
\begin{equation}
    S(\omega) =\int_{\xi_{\rm min}}^{\xi_{\rm max}} d\xi P(\xi) S_{\xi}(\omega).
    \label{eq:s_multiple}
\end{equation}
If the switching rates have a distribution that is proportional to the inverse of the rate, that is, $P(\xi) \propto 1/\xi $ within this range,  Eq.(\ref{eq:s_multiple}) yields $S(\omega) \propto 1/\omega$, consistent with the $1/f$ noise spectrum~\cite{decohrtn3}. 

In the remainder of the paper, we consider a bosonic mode coupled to a fixed number of RTN fluctuators $N_f$, each with a switching rate sampled from the probability distribution
\begin{equation}
     P(\xi) =  \frac{1}{\ln(\frac{\xi_{\rm max}}{\xi_{\rm min}})} \frac{1}{\xi}.
     \label{eq:xi_dist}
\end{equation}
We assume a uniform coupling strength for all fluctuators, with the switching rates following the above distribution $P(\xi)$. The Hamiltonian then is~\cite{bcm}
\begin{equation}
    \hat{H}_{1/f}= \epsilon \hat{a}^{\dagger} \hat{a} + \frac{\nu}{\sqrt{N_f}}\sum_{i=1}^{N_f} c_i(t)  \hat{a}^{\dagger} \hat{a},
    \label{eq:hamiltonian_multiple}
\end{equation}
where the $i^{th}$ fluctuator has a switching rate $\xi_i$. 

With a similar change of variables $\tau= \nu t$ and $r= \xi/\nu$, for a general bosonic state, the evolved density matrix can be written as
\begin{equation}
    \hat{\rho}_{1/f}(\tau)= \sum_{m,n= 0}^{\infty} \langle e^{i\Phi_{1/f}(\tau) (m-n)}\rangle_{\Phi_{1/f}(\tau)} \rho_{m,n} \ketbra{m}{n}.
\end{equation}
Here, $\Phi_{1/f}(\tau) = \frac{1}{\sqrt{N_f}} \sum_i \phi_{i,\text{RTN}}(\tau)$ where $\phi_{i,\text{RTN}}(\tau)= \int_0^{\tau} d\tau^{\prime} c_i(\tau^{\prime} ) $. The averaging operation $\langle .\rangle_{\Phi_{1/f}(\tau)}$ can be computed explicitly as follows 
\begin{align}
    \langle e^{i\sum_i \phi_{i,\text{RTN}}(\tau) (m-n)}&\rangle _{\Phi_{1/f}(\tau)} \nonumber \\
    =&\langle e^{i \phi_{1,\text{RTN}}(\tau) (m-n)} e^{i \phi_{2,\text{RTN}}(\tau) (m-n)}....\rangle \nonumber \\ 
    =&(\overline{\langle e^{i \phi_{\text{RTN}}(\tau) (m-n)}\rangle})^{N_f},
\end{align}
where the first averaging $\langle ..\rangle_{\phi_{\text{RTN}}(t)}$ is over different instances of the RTN as before, the second averaging operation $\overline{ \langle ..\rangle}$ is performed over the distribution of switching rates of the fluctuators, and $\phi_{\text{RTN}}(\tau)$ is a function of the ratio $r$. Using the distribution $P(r)$, the dephasing function is expressed as 
\begin{align}
    \langle e^{i\sum_i \phi_{i,\text{RTN}}(\tau) (m-n)}&\rangle =   \nonumber \\
    &\Big(\int_{r_{min}}^{r_{max}} \ dr\ \langle e^{i \phi_{\text{RTN}}(\tau) (m-n)}(r) \rangle P( r) \Big)^{N_f},
    \label{eq:int_multiple}
\end{align}
where $r_{\rm min}$ and $r_{\rm max}$ are the minimum and maximum values of the ratio of switching rate to coupling strength that a single or multiple TLSs can have. 
Substituting the previously derived expression for the dephasing function Eq.\eqref{eq:dephasingfunction}, this integral can be evaluated analytically (see Appendix~\ref{app:multiple_fluctuator_derivation}).

\subsection{Simulation and analytical results for the dynamics of a coherent state}
\label{sec:evolution-numerical-results}

\begin{figure*}
    \centering
\includegraphics[width=0.9\linewidth]{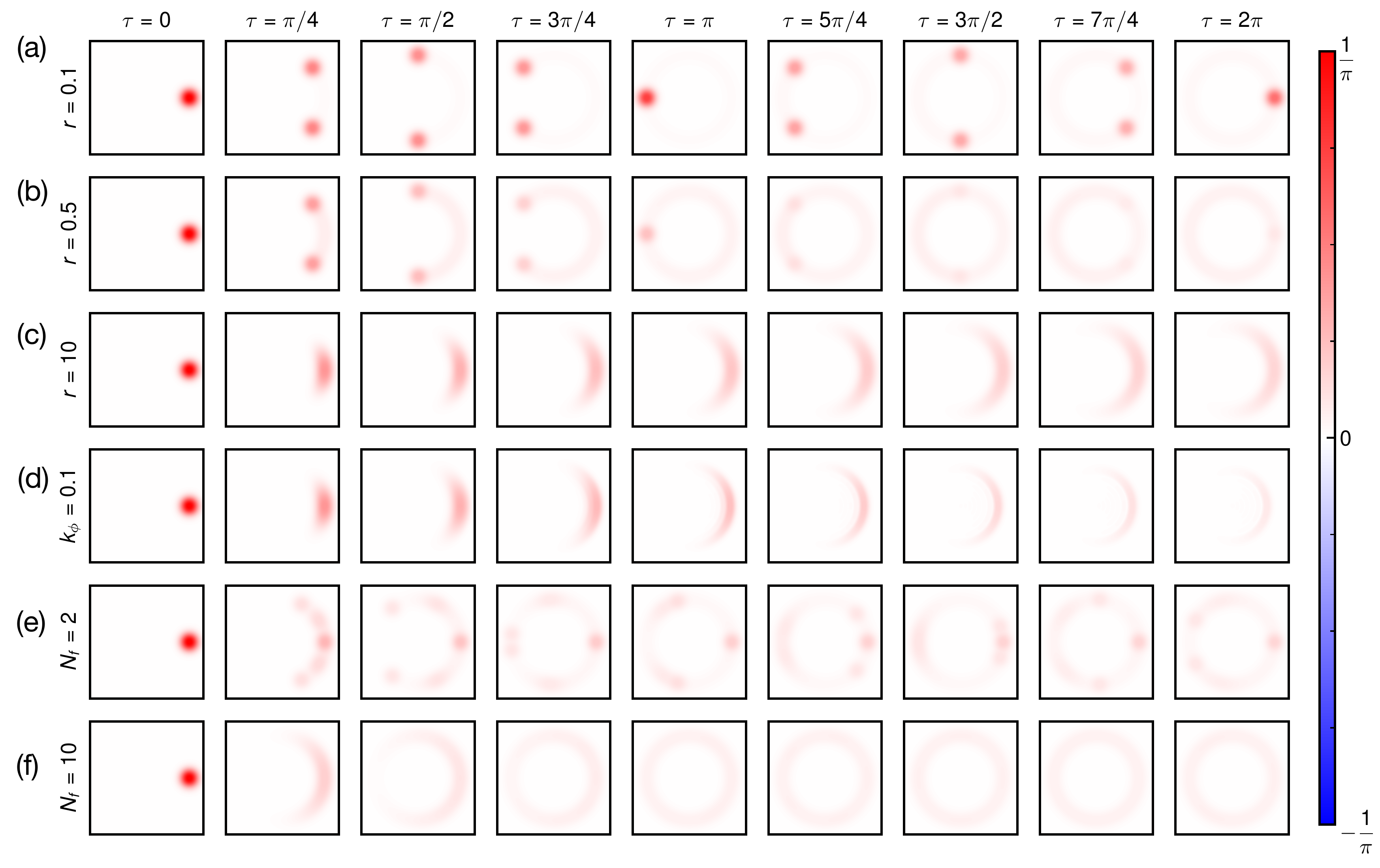}
    \caption{
Evolution of the Wigner function of a coherent state with \(\alpha = 2\) under RTN-induced dephasing for (a) \(r = 0.1\), (b) \(0.5\), and (c) \(10\). For small values of \(r\), the state initially exhibits two distinct lobes, indicative of non-Markovian dynamics, before fully dephasing. As the ratio \(r\) increases, the behavior transitions to one resembling Gaussian dephasing ((d) where dephasing follows a Gaussian decay rate \(k_{\phi}= 0.1\)). Under \(1/f\) noise (e) and (f), similar non-Markovian "blob-like" features appear in the Wigner function for a small number of fluctuators \(N_f\). When \(N_f\) is increased to 10 (f), the behavior becomes similar to that of the Gaussian case.
}
    \label{fig:wigner_function}
\end{figure*}

Before investigating the presence of non-Markovian effects for the above two cases, we explore the evolution of the Wigner function~\cite{Serafini} of the bosonic mode.
The Wigner function of a single-mode bosonic mode $\hat{\rho}$ is given by 
\begin{equation}
    W(q,p)=\frac{1}{2\pi}\int \dd x\;  e^{-ipx} \bra{q+\frac{x}{2}}\hat{\rho} \ket{q-\frac{x}{2}}
\end{equation}
defined over the phase space $(q,p) \in \mathbb{R}^2$. In Fig.~\ref{fig:wigner_function}, we plot evolution of the Wigner function for a coherent state with amplitude $\alpha=4 $ under RTN dephasing with different values of $r$ (Fig.~\ref{fig:wigner_function} (a), (b), (c)), for the case of Gaussian dephasing (Fig.~\ref{fig:wigner_function} (d)) and for $1/f$ noise for $N_f=2, 10$ (Fig.~\ref{fig:wigner_function} (e), (f)). For RTN dephasing with smaller values of $r$, the dominant contribution to the density matrix at early times comes from two ``blob-like'' structures that rotate in opposite directions in the phase space, and that eventually dephase completely at longer times. As $r$ increases, the contribution of states between the blobs becomes larger, and the state dephases over a shorter timescale. In Fig.~\ref{fig:wigner_function} (d), we show the evolution of the same coherent state under a Gaussian dephasing channel. A Gaussian dephasing channel is described by its action on an initial state $\hat{\rho}_0$ as  $\mathcal{N}_{\sigma^2} \hat{\rho}_0 = \int_{-\infty}^{\infty} d\phi~ p(\phi) e^{-i\phi \hat{n}} \hat{\rho}_0 e^{i\phi \hat{n}}$, with dephasing angles $\phi$ drawn from a Gaussian distribution with variance  $\sigma^2 = k_{\phi}\tau$ where $k_{\phi}= k_{\phi}^{\prime}/\nu$ and $k_{\phi}^{\prime}$ is the rate of dephasing. 
The results suggest that as the switching rate of the RTN noise increases, the noise behavior tends to approach the Gaussian (Markovian) limit. This observation is consistent with the understanding that at higher switching rates and longer times, the RTN behaves increasingly like a Markovian channel. In Fig.~\ref{fig:wigner_function} (e) and (f), we show the evolution of the same state under $1/f$ noise with the range of $r$ going from $r_{min}= 10^{-4}$ to $r_{max} = 10^4$ taken from experimental relevant values~\cite{exptls1, exptls4}. For two fluctuators, the evolution of the Wigner function has blob-like behavior showing signs of non-Markovianity; however, for a large number of fluctuators, the evolution once again becomes similar to that of the Gaussian dephasing case (d). 

We now analyze the results of RTN dephasing analytically in both the small- and large $r$ regimes. 
From Eq.~\ref{eq:rho_evolved}, when $r\ll 1$ and on shorter timescales, the formation of the blobs can be understood as follows. Consider the initial state to be a coherent state $|\alpha \rangle$ with $\alpha= \alpha_0 e^{i\theta}$, and the initial density matrix $\hat{\rho}_0 = |\alpha_0, \theta \rangle \langle \alpha_0, \theta | $. The expression for the density matrix in the zeroth order of $r$ and $\tau$ becomes (see Appendix~\ref{app:two_limits})
\begin{equation}
  \hat{\rho}(\tau) = \frac{1}{2} |\alpha_0, \theta+ \tau \rangle \langle \alpha_0, \theta+ \tau | + \frac{1}{2} |\alpha_0, \theta- \tau \rangle \langle \alpha_0, \theta-\tau |,
\end{equation}
where we have dropped the subscript RTN for convenience. 
Thus, at short times, the state consists of two coherent states with phase angles evolving in opposite directions. This explains the blob-like behavior observed in the phase space, as shown in Fig.~\ref{fig:wigner_function}(a). 

In the other limit where $r\gg 1, m,n$, we expand $\Omega$ up to $O(1/r)$, and the evolved density matrix in the Fock basis takes the form
\begin{equation}
    \hat{\rho}(t)= \sum_{m,n= 0}^{\infty}  e^{- \frac{1}{2}\frac{(m-n)^2\tau}{r}}  \rho_{m,n} \ket{m} \bra{n}.
    \label{eq:gaussian_limit}
\end{equation}
For a Gaussian dephasing channel, the evolved density matrix of a state written in Fock basis has the form $\hat{\rho}_G(t)=\sum_{m,n=0}^{\infty}  e^{-\frac{1}{2}(m-n)^2\sigma^2} \rho_{m,n} \ket{m} \bra{n}$ (See Appendix~\ref{app:two_limits}).
By comparing this with the result above, we find that the RTN dephased state in the large 
$r$ limit is similar to that of a Gaussian dephasing channel with an effective dephasing rate $k_{\phi}^{\prime}= \nu^2/\xi$. \\

\section{Quantifying non-Markovianity}
\label{sec:quantifying_N_blp}
 Among the different notions of quantum Markovianity specifically devised to deal with continuous variables~\cite{fidelity,Torre2018,Richter2022, Gaussian_channels}, we will focus here on the BLP  measure~\cite{tracedistance, Breuer2016} based on the trace distance between two states. 
The trace distance between the states $\hat{\rho}_1$ and $\hat{\rho}_2$ is defined as~\cite{tracedistance}
\begin{equation}
    D(\hat{\rho}_{1}, \hat{\rho}_{2})= \frac{1}{2}\Tr |\hat{\rho}_{1}-\hat{\rho}_{2}|= \frac{1}{2}\Tr \sqrt{(\hat{\rho}_{1}-\hat{\rho}_{2})^{\dagger}(\hat{\rho}_{1}-\hat{\rho}_{2})}.
    \label{eq:tracedistance}
\end{equation}
In addition to being a metric, the trace distance has a clear physical interpretation, as it directly quantifies the distinguishability between two states~\cite{tracedistance}. The key idea behind this measure is that a Markovian process, which involves continuous information flow from the system to the environment, tends to reduce the distinguishability between any two states over time. In contrast, for non-Markovian processes, which allow for information backflow, the distinguishability between states can increase at certain time intervals. These revivals in the trace distance over time are a signature of non-Markovianity~\cite{tracedistance, tracedistance1, tracedistance2, tracedistance3}. Given the rate of change of the trace distance,
\begin{equation}
    \sigma(t, \hat{\rho}_{1,2} (t)) = \frac{d}{dt} D(\hat{\rho}_{1}(t), \hat{\rho}_{2}(t)),
\end{equation}
Markovian processes are defined by the condition $\sigma(t, \hat{\rho}_{1,2} (t)) \leq 0$ for all times $t$ and for all pairs of initial states $\hat{\rho}_1(0)$  and $\hat{\rho}_2(0)$ i.e., the trace distance is a monotonically decreasing function of time. Instead, for a non-Markovian process, this quantity becomes greater than zero for some time intervals,
indicating a backflow of information to the system. 
To quantify this total amount of information flowing from the environment back to the system during the dynamics, the non-Markovianity measure is defined as

\begin{equation}
\mathcal{N}_{\text{BLP}} = \underset{\hat{\rho}_1(0), \hat{\rho}_2(0)}{\max}\int_{\dot{\sigma}>0}  dt \ \sigma(t, \hat{\rho}_{1,2}(0)).
 \label{eq:nblp}
\end{equation}
By construction, the value of this measure $\mathcal{N}_{\text{BLP}}$ is zero for Markovian processes and is greater than zero for non-Markovian processes. 

Notably, in Appendix~\ref{app:wigner_neg} we have also cross-checked our results with another measure of non-Markovianity related to the Wigner negativity of a state. We see that the two approaches agree qualitatively; hence, we focus on the trace-distance-based measure for the rest of the analysis. 
\subsection{$\mathcal{N}_{BLP}$ for Gaussian states}
\label{sec:gaussian_and_gaussian}

In this subsection, we characterize the non-Markovianity as measured by BLP for Gaussian states, for both RTN and $1/f$ noise.

\subsubsection{Gaussian states evolving under RTN noise}
\label{sec:RTN-Gaussian}

To investigate the non-Markovian effects induced by RTN, we analyze the evolution of the trace distance between two initial bosonic modes, focusing on the occurrence of revivals. The $\mathcal{N}_{\text{BLP}}$ measure is then evaluated by maximizing the summation of revivals over all possible bosonic states. However, this maximization is computationally prohibitive due to the unbounded domain of several parameters involved for the infinite set of CV states. We thus begin by confining our investigation to a broad class of Gaussian states.
A general Gaussian state is expressed as~\cite{Serafini}
\begin{equation}
    \hat{\rho}^G = \hat{D}(\alpha) \hat{S}(\beta) \hat{\nu}_{th}(\tilde{N}) \hat{S}^{\dagger}(\beta)\hat{D}^{\dagger}(\alpha),
    \label{eq:general_gaussian}
\end{equation}
where $\hat{D}(\alpha)= \exp(\alpha \hat{a}^{\dagger}- \alpha^* \hat{a})$ is the displacement operator, $\hat{S}(\beta)= \exp(\frac{1}{2}(\beta \hat{a}^{\dagger^2} - \beta^* \hat{a}^{2}))
$ is the squeezing operator and $\hat{\nu}_{th}= \tilde{N}^{\hat{a}^{\dagger}\hat{a}}/(\tilde{N}+1)^{(\hat{a}^{\dagger}\hat{a}+1)}$ is the thermal equilibrium state with $\tilde{N}$ average photon number.
The parameters $\alpha$ and $\beta$ are complex numbers characterised by their amplitude and phase $(\alpha_0, \theta)$ and $\alpha =\alpha_0 e^{i \theta}$ and $\beta=\beta_0 e^{i\gamma}$ respectively. The parameter $\alpha$ determines the position of the state in the phase space, and $\beta$ specifies the magnitude and direction of squeezing. 
To evaluate $\mathcal{N}_{\text{BLP}}$, the optimization spans ten real parameters from the initial two Gaussian states: $(\alpha_{0_1}, \alpha_{0_2}, \theta_1, \theta_2, \beta_{0_1}, \beta_{0_2}, \gamma_1, \gamma_2, \tilde{N}_1, \tilde{N}_2)$.
This procedure is computationally intensive, as the domains of some of the above parameters are unbounded. However, we see that not all ten parameters are relevant to the maximization procedure. 
We compute the sum of trace distance revivals for a specific pair of states and define this quantity as $\mathcal{N}_{\text{BLP}}^{*}$ (denoted with an asterisk):

\begin{equation}
    \mathcal{N}_{\text{BLP}}^{*} (\hat{\rho}_1(0), \hat{\rho}_2(0)) = \int_{\sigma>0}  \dd t \ \sigma(t, \hat{\rho}_{1,2}(0)).
 \label{eq:nblp_state}
\end{equation}
This quantity is still dependent on the chosen initial states. 

As detailed in Appendix~\ref{app:coherent_state_maximisation}, we begin by optimizing the bounded parameters $\theta$ and $\gamma$, and we see that we obtain the maximum value of $\mathcal{N}_{\text{BLP}}^{*}$ when $|\theta_1 - \theta_2| = \pi$ and $\gamma_1 = \gamma_2 = 2\theta_1$. All the combinations of $\theta_1$ and $\theta_2$ and values of $\gamma_1$ and $\gamma_2$ that satisfy the above condition have the same value of $\mathcal{N}_{\text{BLP}}^{*}$ owing to the dephasing nature of the random telegraph noise. Now we consider the parameters $\alpha_{0}, \beta_{0}$ and $\tilde{N}$ for a given state, and we see that $\mathcal{N}_{\text{BLP}}^{*}$ is maximised when $\alpha_{0_1} = \alpha_{0_2}= \alpha_0$, $\beta_{0_1} = \beta_{0_2} = \beta_0 $ and $\tilde{N}_1= \tilde{N}_2= \tilde{N}$. This reduces the optimization to three parameters: $\alpha_0$, $\beta_0$, and $\tilde{N}$. Upon further analysis, we observe in Appendix~\ref{app:coherent_state_maximisation} that neither the average number of quanta nor the squeezing parameter increases the measure of non-Markovianity. Thus $\mathcal{N}_{\text{BLP}}^{*}$ is maximised for a given value of $\alpha_0$ with $\beta_0=0$ and $\tilde{N}=0$. 

\begin{figure}
    \centering
    \includegraphics[width=0.9\linewidth]{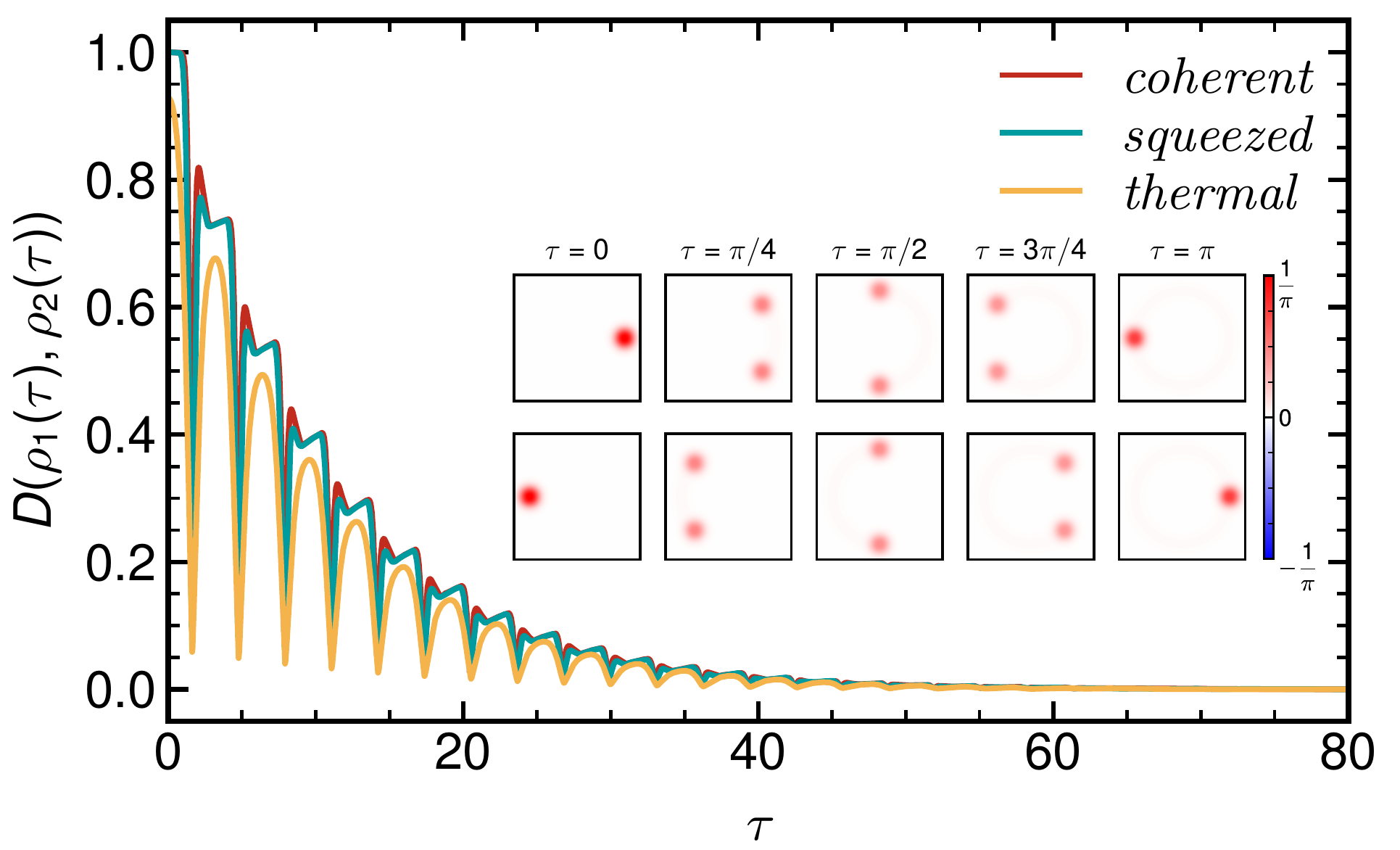}
    \caption{Evolution of the trace distance between pairs of quantum states: coherent states with \(\alpha_1 = -\alpha_2 = 4\) (red), squeezed states with \(\beta_1 = \beta_2 = 0.4\) (blue), and thermal states with \(\tilde{N}_1 = \tilde{N}_2 = 2\) (yellow). We see that squeezing and thermal fluctuations lead to a suppression of oscillation amplitudes in the trace distance dynamics, thereby reducing the effective non-Markovianity measure \(\mathcal{N}_{\mathrm{BLP}}^{*}\).}
    \label{fig:gaussian}
\end{figure}

In Fig.~\ref{fig:gaussian}, we plot the evolution of trace distance as a function of $\tau$ for $r=0.1$ for a pair of coherent states with $\alpha_0 = 4$ and phase difference of $\pi$, squeezed states with $\beta_0=0.4$ and thermal states with $\tilde{N}=2$ in the non-Markovian regime.  While oscillations indicating non-Markovian behavior are present in the cases of squeezed and thermal states, their amplitudes are reduced relative to those of coherent states. This results in the measure of non-Markovianity $\mathcal{N}_{\text{BLP}}^{*}$ being maximum for the coherent states among the three cases. We further illustrate the Wigner evolution of two initial states $\alpha_0 =4$ and phase difference $\pi$. The periodic oscillations in the trace distance can be qualitatively understood from this visualization. When $\tau$ equals odd multiples of $\pi/2$, significant overlap between the initial blobs causes $\hat{\rho}_1 - \hat{\rho}_2 \rightarrow 0 $, resulting in a dip in the trace distance at those times.

Finally, we examine the maximum value of $\mathcal{N}_{\text{BLP}}^{*}$ as a function of $\alpha_0$ for a fixed $r$. We find that a specific value of  $\alpha_{0}$, that we call $\alpha_{0_{\rm max}}$, maximises the measure, and this value increases with $r$. The plots $\mathcal{N}_{\text{BLP}}^{*}$ as a function of $\alpha_0$ for different values of $r$ are given in Fig.~\ref{fig:main_results2}(a). For the values of $r$ considered, we see from our numerical calculations that for the class of Gaussian states represented by Eq.(\ref{eq:general_gaussian}), a pair of coherent states with a particular amplitude  $\alpha_{0_{\rm max}}$ maximizes the non-Markovian measure. Although the analytical relationship between  $\mathcal{N}_{\text{BLP}}^{*}$ and $\alpha_{0_{\rm max}}$ remains elusive, our numerical results indicate that the value of $\alpha_{0_{\rm max}}$ increases with increase in the value of the ratio $r$, making it computationally challenging for higher photon numbers. We also observe, as expected, that the value of $\mathcal{N}_{\text{BLP}}^{*}$ increases when $r$ decreases, since this corresponds to going into a stronger non-Markovian regime. 

\subsubsection{Gaussian states evolving under $1/f$ noise}
\label{sec:1/f}

 In Fig.~\ref{fig:main_results2}(c), we plot the numerical results for the simulation of $1/f$ noise for a pair of coherent states with different values of $\alpha$. We have fixed the range of values of $r$ over which the integration in Eq.(\ref{eq:int_multiple}) is performed to be $[10^{-4}, 10^4]$. However, we see that values above $r=1$ do not contribute significantly to the non-Markovianity. The results show that as $N_f$ increases, the oscillation amplitude of the trace distance diminishes (See Fig.~\ref{fig:1/f}(a) in Appendix~\ref{app:1/f}). Consequently, the non-Markovianity measure exponentially decreases with increasing $N_f$. 
 Experimental studies often report a small number of fluctuators coupling to the system of interest~\cite{exptls1, exptls4}. Therefore, in realistic scenarios where $N_f \lessapprox 10$, significant non-Markovian behavior is still expected, as reflected in our results.

\subsection{$N_{BLP}$ for non-Gaussian states}
\label{sec:non_gaussian}
While we cannot establish a bound on non-Markovianity for all bosonic states, it is insightful to explore non-Markovian effects within a subclass of non-Gaussian states known as rotation symmetric bosonic (RSB) codes~\cite{qec3}, which have exhibited robust error-correcting properties~\cite{arne, timo}.
A code exhibits discrete 
$ N$-fold rotational symmetry if every state in its code space is an eigenstate of the discrete rotation operator~\cite{arne}
\begin{equation}
    \hat{R}_N = e^{i \frac{2\pi}{N} \hat{n}},
\end{equation}
with eigenvalue $+1$. We also define a logical $\hat{Z}_N$ operator as the square root of $\hat{R}_N$ which also preserves the rotational symmetry, 
\begin{equation}
    \hat{Z}_N= \hat{R}_{2N} = e^{i \frac{\pi}{N} \hat{n}}.
\end{equation}
As such, $N$-order RSB code is characterised by $\pm$ eigenvalue of the logical $\hat{Z}_N$ on the codespace, represented by the projector $\hat{\Pi}_{\rm code}= \ketbra{0}{0}_{\rm code} + \ketbra{1}{1}_{\rm code} $. 

Given that the logical codewords $\ket{0}_N$ and $\ket{1}_N$ satisfy the eigenvalue condition $\hat{Z}_N \ket{j}_N = (-1)^{j} \ket{j}_N$ for $j \in {0,1}$, they can be constructed from discrete rotated superpositions of a primitive state $\ket{\Theta}$ such that 
\begin{eqnarray}
    \ket{0}_{N,\Theta} &=& \frac{1}{\sqrt{N_0}}\sum_{m=0}^{2N-1} e^{i\frac{m\pi}{N}\hat{n}}\ket{\Theta}\\
    \ket{1}_{N,\Theta} &=& \frac{1}{\sqrt{N_1}}\sum_{m=0}^{2N-1} (-1)^me^{i\frac{m\pi}{N}\hat{n}}\ket{\Theta},
    \label{eq:rsb1}
\end{eqnarray}
where ${N}_i $ are the normalisation constants.
By expressing the primitive state in Fock basis as $\ket{\Theta}= \sum_n c_n \ket{n}$, the codewords can be written as:
\begin{eqnarray}
        \ket{0}_N &=& \sum_k f_{2kN} \ket{2kN}\\
    \ket{1}_N &=& \sum_k f_{(2k+1)N} \ket{(2k+1)N},
    \label{eq:rsb2}
\end{eqnarray}
where $f_{kN}$ are coefficients determined by the primitive state.
In the dual basis, the codewords transform into
\begin{eqnarray}
    \ket{\pm}_N &=& \frac{1}{\sqrt{2}}(\ket{0}_N + \ket{1}_N) \nonumber \\
    &=&  \frac{1}{\sqrt{2}} \sum_k^{\infty} (\pm1)^k f_{kN} \ket{kN}.
    \label{eq:rsb3}
\end{eqnarray}
This implies that, for the codewords to be well-defined, the primitive state must have non-zero support on at least one of the $\ket{2kN}$ and one of the $\ket{2kN+1}$ states. 
In this work, we focus on two specific categories of rotation-symmetric bosonic codes: binomial codes and cat codes. The performance of these codes has been extensively studied in~\cite{arne} and~\cite{timo}, % \commg{Timo: can you please add further relevant references?}, 
demonstrating promising results for state recovery using error correction circuits under the effects of photon loss and Gaussian dephasing. 
Additionally, Refs.~\cite{ouyang_trade-offs_2020, endo_quantum_2022, optimal} study RSB codes under various other noise models.

For cat codes, the computational codewords $\ket{0_N}$ and $\ket{1_N}$ are constructed by taking the primitive state in Eq.(\ref{eq:rsb1}) as the coherent states, $\ket{\Theta}= \ket{\alpha}= e^{-|\alpha|^2/2} \sum_n \alpha^n/\sqrt{n!} \ket{n}$. The rotated superpositions of these states form the codewords as 
\begin{eqnarray}
    \ket{0}_{N} &=& \frac{1}{\sqrt{N_0}}\sum_{m=0}^{2N-1} \ket{\alpha e^{i\frac{m\pi}{N}}}\\
    \ket{1}_{N} &=& \frac{1}{\sqrt{N_1}}\sum_{m=0}^{2N-1} (-1)^m \ket{\alpha e^{i\frac{m\pi}{N}}}.
    \label{eq:cat}
\end{eqnarray}

For the class of binomial codes, the codewords have the following form, with the coefficients $f_{kN}$ given by binomial distribution~\cite{arne} \footnote{The definition of the codeword $\ket{1}_{N, K}$ has a typo in the~\cite{arne}, which has been corrected here.}:
\begin{eqnarray}
     \ket{0}_{N,K}&=& \sum_{k=0}^{\lfloor K/2 \rfloor} \sqrt{\frac{1}{2^{K-1}}\binom K {2k}} \ket{2kN} \\
    \ket{1}_{N,K}&=& \sum_{k=0}^{\lceil K/2 -1 \rceil} \sqrt{\frac{1}{2^{K-1}}\binom K {2k+1} } \ket{(2k+1)N}.
    \label{eq:binom}
\end{eqnarray}
These codes can correct against photon losses, photon gains, and dephasing. The parameter $K$ is the truncation limit and determines the maximum order up to which the code can correct for the above-mentioned errors. 

%\subsubsection{Non-Markovianity for bosonic codewords evolving under RTN noise}
%\label{sec:RTN-non-Gaussian}

\begin{figure}
\centering
\includegraphics[width=\linewidth]{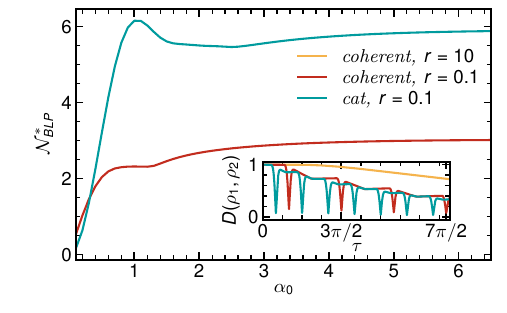}
\caption{
 Non-Markovianity measure $\mathcal{N}_{\mathrm{BLP}}^{*}$ for \(r = 0.1\) as a function of the parameter \(\alpha_0\), evaluated for a pair of coherent states with $\alpha_0=2$ and a phase difference of $\pi$, (shown in red) and the codewords of cat states \(\ket{+}_{2, \alpha}\) and \(\ket{-}_{2, \alpha}\), (shown in blue). The measure is almost twice in the latter case. In the inset, we show that for the cat states, the oscillation period of the trace distance between the codewords is half that of a pair of coherent states with the same \(\alpha\), resulting in doubly frequent revivals and a non-Markovianity measure nearly twice that of the coherent states. Additionally, in the inset, we show in yellow how the oscillations in the trace distance disappear in the Markovian limit of $r=10$. 
}
\label{fig:cat}
\end{figure}

The non-Markovian effects on the evolution of the binomial codewords in the presence of  RTN as a function of the noise strength and covering both Markovian and non-Markovian regimes, are provided in Fig.~\ref{fig:main_results2}(a) and (b). We observe that the period of oscillations depends linearly on the order of rotation symmetry $N$, however, the peaks of the oscillations only broaden with an increase in $K$.
Here, we evaluate the non-Markovian measure $\mathcal{N}_{BLP}^{*}$ for the corresponding cases of cat states. 
In Fig.~\ref{fig:cat}(a) we see $\mathcal{N}_{BLP}^{*}$ for the case of cat states ($\ket{+}_{2,\alpha}$ and$\ket{-}_{2,\alpha}$) compared to the coherent states ($\ket{+\alpha}$ and $\ket{-\alpha}$) for the same value of $\alpha$ doubles in magnitude. This directly reflects the doubling of the frequency of the trace distance due to the underlying $N=2$ symmetry of the code. For similar reasons, for the binomial code, in Fig.~\ref{fig:main_results2}(b), we notice that the measure linearly increases with the order of symmetry of the code $N$.  As a further remark, we show in Appendix~\ref{app:nblp_non_gaussian} that for a general state in the span of Fock states $\ket{0}$ and $\ket{l}$ where $l \in \mathbb{N}$, the value of non-Markovian measure goes to infinity as $l \rightarrow \infty $. 
An analysis of the effects of $1/f$ noise on the codewords of the binomial states ($\ket{+}_{1,2}$ and $\ket{-}_{1,2}$) is provided in Appendix~\ref{app:1/f} where we again see that the non-Markovianity measure approaches zero as $N_f \gtrsim 10$. The behavior is similar to the case of the coherent state, and thus we also expect an exponential decay of $\mathcal{N}_{BLP}^{*}$ as a function of $N_f$ even for the case of binomial codes.

\section{Performance of RSB Codes Under RTN Dephasing and $1/f$ Noise}
\label{sec:performance}

 The RSB codes introduced in the previous section have been shown to perform particularly well under loss and Gaussian dephasing \cite{arne, timo} with a Knill-type error correction protocol. 
 In this section, we proceed to investigate their performance under RTN dephasing and $1/f$ noise along with a loss channel. 
\subsection{Knill Error Correction Circuit and Recovery Map}
\label{sec:circuit_details}
% Although the practical implementation of quantum computing architectures is limited by decoherence processes, the theory of fault-tolerant quantum computation guarantees reliable operation provided that physical error rates remain below a certain threshold. 
% In this direction, bosonic codes offer a promising path forward, exploiting the infinite-dimensional Hilbert space of a single mode to provide hardware-efficient error correction alternatives to conventional multi-qubit codes~\cite{qec1, qec2, qec3}. 
% Notably, recent experiments have demonstrated that bosonic codes can surpass the break-even point, where error-corrected qubits outperform their unencoded counterparts~\cite{exp_breakeven, sivak2023real, ni2023beating}.

Knill-type teleportation-based error correction circuits adapted for rotation-symmetric bosonic (RSB) codes have been numerically analyzed in prior works~\cite{arne, timo}. 
The results from \cite{arne} for RSB codes find high average fidelities under idealized measurement scenarios, and ~\cite{timo} extends the work to physically relevant measurement schemes. 
However, their analysis has predominantly focused on noise models incorporating loss and Gaussian dephasing. 
Here we extend this investigation to more realistic dephasing environments characterized by random telegraph noise (RTN) and $1/f$ noise. 
We assume idealized conditions for state preparation, gate operations, and auxiliary qubits~\cite{arne,timo} to isolate the effects of these dephasing channels.

We review the Knill-EC protocol for RSB codes, then evaluate the performance under a noise channel $\mathscr{N}$ incorporating photon loss and RTN dephasing. 
Our analysis focuses on how code performance evolves with varying RTN parameters, transitioning from the Markovian to non-Markovian regimes.

The Knill-EC circuit for RSB codes, as introduced in~\cite{arne}, includes the preparation of $\ket{+}_N$ states, CROT gates (controlled rotations) that act as logical controlled-Z gates on encoded codewords, and measurements to distinguish logical codewords in the dual basis.
The protocol performs two consecutive one-bit teleportations to transfer the encoded state onto a fresh auxiliary mode, mitigating accumulated noise.

\begin{figure}
    \centering
    \includegraphics[width=\linewidth]{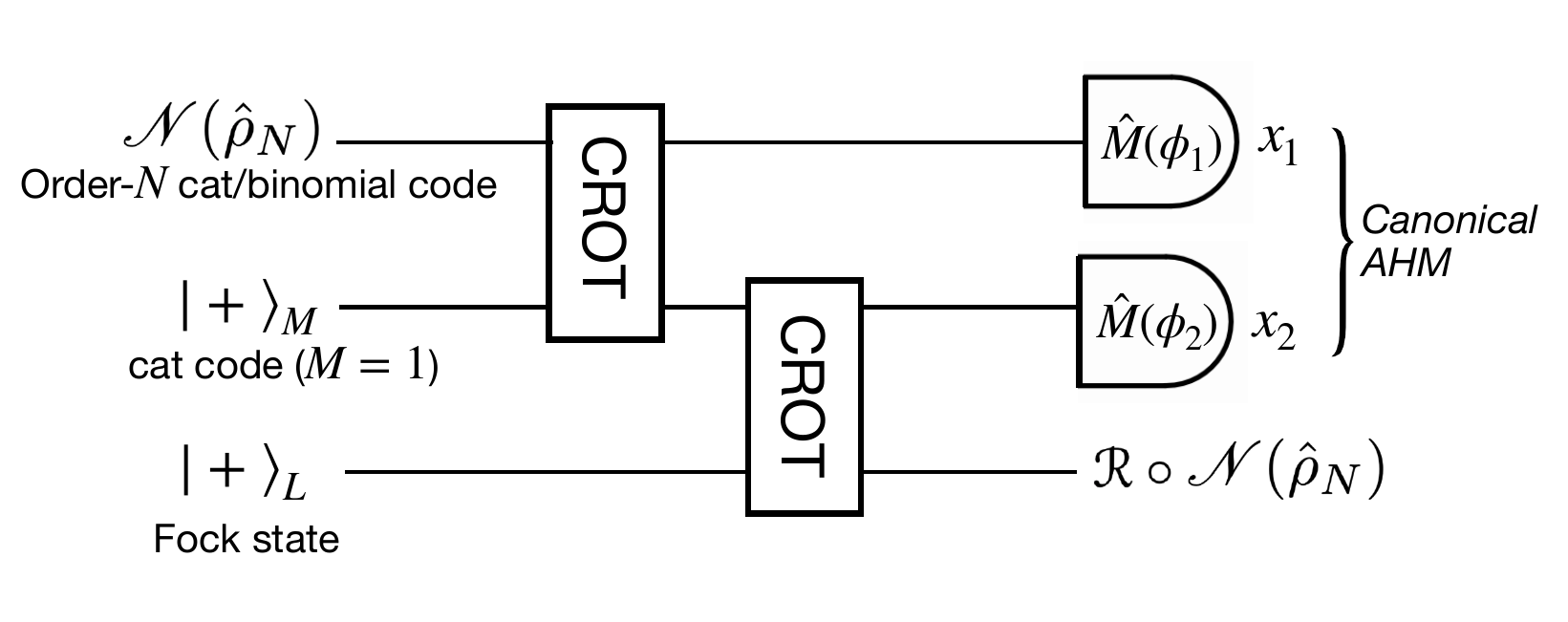}
    \caption{Teleportation-based Knill-EC circuit for bosonic codes. }
    \label{fig:circuit_ec}
\end{figure}

The data rail holds the initial logical state $\hat{\rho}_N = \sum_{i,j=0}^1 \rho_{ij} \ket{i}_N\bra{j}_N$ encoded in an order-$N$ rotation code. 
Auxillary modes are prepared in $\ket{+}_M$ and $\ket{+}_L$ for sequential teleportations. 
The data mode is first subjected to a combined noise channel, with photon loss and RTN dephasing. 
The photon loss channel is governed by the Lindblad master equation:
\begin{equation}
    \frac{d}{dt}\hat{\rho}(t) = \kappa \mathcal{D}[\hat{a}] \hat{\rho}(t),
\end{equation}
where $\kappa$ is the photon loss rate and $\mathcal{D}[\hat{L}] \hat{\rho} = \hat{L} \hat{\rho} \hat{L}^{\dagger} - \frac{1}{2} \{ \hat{L}^{\dagger} \hat{L}, \hat{\rho} \}$ with $\{\hat{A}, \hat{B}\}$ being the anti-commutator of operators $\hat{A}$ and $\hat{B}$. 
The corresponding Kraus operators are:
\begin{equation}
    \hat{A}_k = \frac{(1 - e^{-\kappa t})^{k/2}}{\sqrt{k!}} e^{-\kappa \hat{n} t / 2} \hat{a}^k,
\end{equation}
with $\hat{a}$ the annihilation operator and $\hat{n} = \hat{a}^{\dagger}\hat{a}$ the number operator.
For RTN dephasing, we numerically obtain Kraus operators from the channel's superoperator formulation (Eq.~\ref{eq:rho_evolved} can be easily recast into superoperator form).

The CROT gate is given by
\begin{equation}
    \text{CROT}_{NM} = e^{i (\pi / NM) \hat{n}_a \otimes \hat{n}_b},
\end{equation}
and serves as a logical controlled-Z gate, acting on the codewords as $\text{CROT}_{NM}\ket{i}_N \otimes \ket{j}_M = (-1)^{ij} \ket{i}_N \otimes \ket{j}_M$. 
The number-shift errors in one mode propagate as rotation errors in the other, and we see that this feature is crucial to obtaining the recovery map for the noisy state (Appendix~\ref{app:circuit_derivation}).

To distinguish logical codewords in a dual basis, we employ a phase-sensitive POVM~\cite{measurement1}:
\begin{equation}
    \hat{M}(\phi) = \frac{1}{2\pi} \sum_{m,n=0}^{\infty} e^{i\phi(m-n)} B_{mn} \ketbra{m}{n},
    \label{eq:povm}
\end{equation}
where $B$ is a positive semi-definite Hermitian matrix and $\phi \in [0, 2\pi)$. We primarily use canonical phase measurements for which $B_{mn}=1$~\cite{measurement1}.
However, we also evaluate the performance under adaptive homodyne detection~\cite{measurement2, measurement3}, an improved physically implementable measurement scheme.

The post-measurement recovery map, summed over outcomes $x_1, x_2$, is given by:
\begin{equation}
    \mathcal{R} \circ \mathscr{N}(\hat{\rho}_N) = \frac{1}{4} \sum_{x_1,x_2} \hat{P}_{i^{*}(\vec{x})}^{\dagger} \left[ \sum_{i,j=0}^{3} c_{ij}(\vec{x})  \hat{P}_i \hat{\rho}_L \hat{P}_j^{\dagger} \right] \hat{P}_{i^{*}(\vec{x})},
    \label{eq:recovery_map}
\end{equation}
where $\hat{P}_i \in \{\mathds{I}, X, Z, XZ\}$ are logical Pauli operators on the final order-$L$ code state $\hat{\rho}_L$. The decoder selects the most likely Pauli correction $i^{*}$ via:
\begin{equation}
    i^{*} = \arg\max_{i} \Tr[\hat{M}_{x_1} \otimes \hat{M}_{x_2} \hat{\mathcal{U}}_c \circ \mathscr{N} \circ \hat{\mathcal{U}}_c^{\dagger} \ket{i} \bra{i}],
\end{equation}
with $\hat{\mathcal{U}}_c = \text{CROT}_{NM} \circ \text{CROT}_{NM}^{\dagger}$ and basis states $\ket{i} = H\ket{a}_N \otimes H\ket{b}_M$. The complete derivation is given in Appendix~\ref{app:circuit_derivation}.

We compute the average gate fidelity~\cite{av_gate_fid} $ \mathcal{F}_{\text{EC}}$ of the resulting output state to assess the performance of these codes. The average gate fidelity of a quantum channel $\mathcal{E}$ is defined by
\begin{align}
    \mathcal{F}(\mathcal{E}) = \int \dd\psi\;\bra{\psi}\mathcal{E}(\ketbra{\psi}{\psi})\ket{\psi},
\end{align}
where $\dd\psi$ is a uniform measure on the state space, normalized to unity: $\int \dd\psi = 1$.

\subsection{Numerical Results}
\label{sec:numerical_results}
In this subsection, we present the numerical results of the performance of the error-correction protocol, obtained by calculating the average gate fidelity~\cite{av_gate_fid}, $\mathcal{F}_{\text{EC}}$ of the output quantum channel obtained after applying the most likely Pauli correction. We further analyze this as a function of various parameters of the noise model and the input encoding state.

\begin{figure*}
    \centering
    \includegraphics[width=\linewidth]{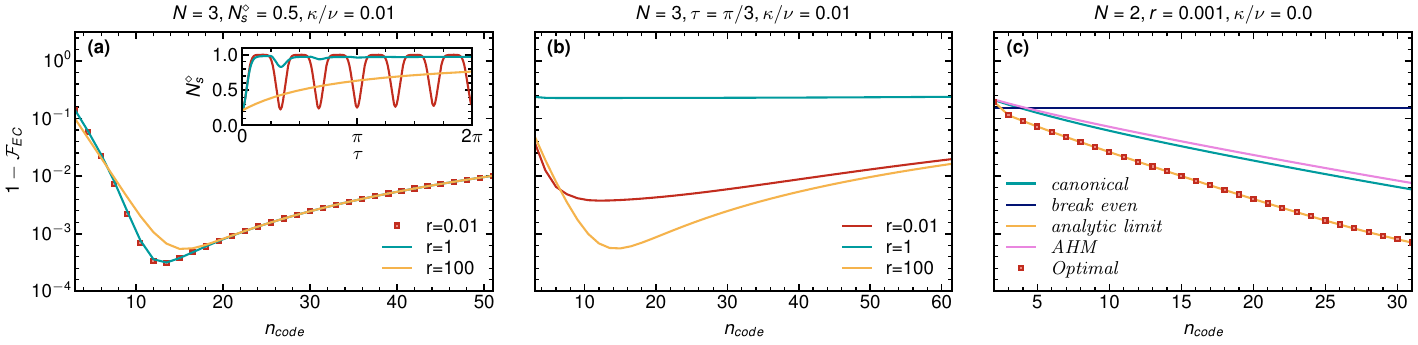}
    \caption{
(a) Infidelity of the error-corrected state recovered using the Knill-EC circuit, shown for various values of the parameter \(r\), ranging from the Markovian regime (\(r = 100\)) to the non-Markovian regime (\(r = 0.01\)), in the presence of an additional photon loss channel with \(\kappa/\nu = 0.01\) as a function of average photon number of the binomial code. The noise strength of different $r$ and for each value of $K$ has been fixed to be $N_s^{\diamond}=0.5$. The inset shows the oscillatory behavior of the noise strength with $\tau$ for $K=8$ (which has the least infidelity for $N=3$). We see that the codes perform significantly well and are qualitatively similar for all values of $r$ at a given noise strength. However, in (b), for a fixed $\tau= \pi/3$, since the noise strength is the highest for $r=1$, the code performs worse for this case. The performance is therefore sensitive to $\tau$ for a given $r$. In the non-Markovian regime, we can expect good performance even at larger but specific values of $\tau$ due to the oscillatory feature of the noise. From the inset of (a), we can see that there are periodically occurring times for which the noise strength becomes small, leading to better performance for those values of $\tau$. 
(c) Infidelity under pure RTN dephasing (no loss) in the strongly non-Markovian regime (\(r = 0.001\)) for \(N = 2\) and \(\tau = 1\). The performance of the Knill-EC circuit (with canonical phase measurements) is compared with the analytically computed fidelity limit, the break-even threshold, optimal recovery map, by using adaptive homodyne detection measurements.
}
\label{fig:fidelity}
\end{figure*}

\begin{figure}
    \centering
    \includegraphics[width=0.8\linewidth]{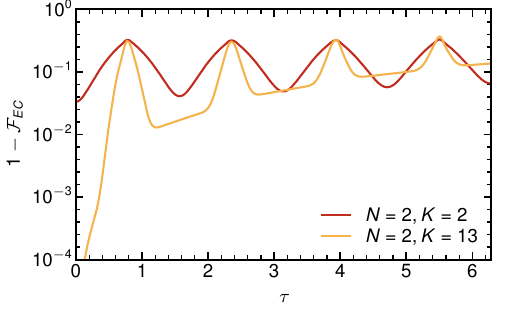}
    \caption{Infidelity from the Knill EC circuit for a binomial code with the given parameters in the non-Markovian limit of $r= 0.1$ with a loss channel of strenght $\kappa/\nu= 0.01$. We notice that the frequency of oscillations depends only on the order of symmetry $N$. For higher values of $K$, we see that the width of the oscillations increases, resulting in longer time intervals with increased performance.}
    \label{fig:fidelity_time}
\end{figure}

In Fig.~\ref{fig:fidelity}(a) and (b), we plot the infidelity ($1 - \mathcal{F}_{\text{EC}}$) as a function of the average photon number of the code $n_{\text{code}}$ for a fixed loss rate $\kappa_{\phi} = 0.01$, various values of $r$ and for (a) fixed noise strength $N_s^{\diamond}$ and various values of $\tau$, and (b) fixed value of $\tau$ and various values of $N_s^{\diamond}$. 
The average photon number is calculated as 
\[
n_{\text{code}} = \frac{1}{2} \mathrm{Tr} \left[ \left( \ketbra{0}_N + \ketbra{1}_N \right) \hat{n} \right],
\] 
and for binomial codes, we have $n_{\text{code}}= \frac{1}{2}NK$. 
We characterize the noise strength in a channel by evaluating the diamond norm of the channel after being projected to the codespace. Defining the projector operator that takes a state from Fock basis $\{\ket{0}, \ket{1}\}$ to the codespace of an RSB code spanned by $\{\ket{0}_N, \ket{1}_N$\} where $N$ is the order of rotation symmetry of the code, we have 
\begin{equation}
    \hat{S}= \ket{0}_L \bra{0} + \ket{1}_L\bra{1}.
\end{equation}
The superoperator form $\mathcal{E}$ of the noise channel map that arises for loss, RTN, or $1/f$ dephasing can then be projected onto the codespace as 
\begin{equation}
    \mathcal{E}^{\prime} = \hat{\tilde{S}}^{\dagger} \mathcal{E} \hat{\tilde{S}},
\end{equation}
where $\hat{\tilde{S}}$ is the superoperator form of $\hat{S}$. We then evaluate the diamond norm $||\mathcal{E}^{\prime}||^{\diamond}$~\cite{diamond1, diamond2, diamond3} of this channel numerically and define noise strength as
\begin{equation}
    N_s^{\diamond}= 1-||\mathcal{E}^{\prime}||^{\diamond}.
    \label{eq:Ns}
\end{equation}
The identity channel will give $||\mathcal{E}^{\prime}||^{\diamond}=1$.
The inset of Fig.~\ref{fig:fidelity}(a) shows the behavior of the noise strength as a function of $\tau$. We see that due to the non-Markovian nature of the noise, for $r=0.01$, we have oscillations in the noise which fade away as we increase $r$. Similarly, we observe that the average gate fidelity of the circuit $\mathcal{F}_{\text{EC}}$ also exhibits oscillations with $\tau$ (Fig.~\ref{fig:fidelity_time}), the period of which depends on $N$ of the code.

To understand the nature of these oscillations, we derive a semi-analytical expression for the average gate fidelity under a general dephasing channel (see Appendix~\ref{app:fidelity_derivation}). Assuming canonical phase measurements (Eq.\eqref{eq:povm}), the fidelity takes the form:
\begin{align*}
    \mathcal{F} &= \frac{1}{2} + \frac{1}{12} \left[ \int\limits_{0}^{2\pi} d\phi_1\, |c_0^N - c_1^N|(\phi_1) + \int\limits_{0}^{2\pi} d\phi_2\, |c_0^M - c_1^M|(\phi_2) \right] \\
    &\quad + \frac{1}{24} \int\limits_{0}^{2\pi} d\phi_1\, |c_0^N - c_1^N|(\phi_1) \int\limits_{0}^{2\pi} d\phi_2\, |c_0^M - c_1^M|(\phi_2),
\end{align*}
where
\[
c_{0/1}^N(\phi_1) = \mathrm{Tr}[\hat{M}(\phi_1) \mathscr{N}(\ket{\pm}\bra{\pm}_N)] \]
\[
c_{0/1}^M(\phi_2) = \mathrm{Tr}[\hat{M}(\phi_2)\ket{\pm}\bra{\pm}_M],
\]
where the auxiliary qubit is assumed to be noiseless.

For $K = 2$ and arbitrary $N$, we find (see Appendix~\ref{app:fidelity_derivation} for derivation)
\[
\int\limits_{0}^{2\pi} d\phi_1\, |c_0^N - c_1^N|(\phi_1) = \frac{4\sqrt{2}}{\pi} |G(N, r, \tau)|,
\]
which is proportional to the modulus of the dephasing function $|G(r, \tau)|$ in Eq.\eqref{eq:dephasingfunction}, oscillating with frequency $\Omega = 2\sqrt{N^2 - r^2}$. For small $r$, $\Omega$ becomes nearly linear in $N$. This trend is visible in numerical simulations as the oscillation period reduces on increasing $N$. Fixing $N$ and increasing $K$ gives a contribution to the fidelity from frequencies higher than $\Omega$, resulting in broader dips in infidelity, and increasingly favorable time windows for correction.
The results in Fig.~\ref{fig:fidelity_time} correspond to binomial codewords in dual basis subjected to a photon loss channel with $r= 0.1$ and $\kappa/\nu = 0.01$ with the code parameters: $N=2, K=2$ (red) and $N=2, K=13$ (yellow). 
In the non-Markovian regime, the frequency of oscillations scales with the code symmetry parameter $N$ for both cases. However, while keeping $N$ and $\kappa/\nu$ fixed, we observe that increasing $K$ modifies the width of the fidelity dips, as shown by the yellow curve in Fig.~\ref{fig:fidelity_time}, thus increasing our chances of remaining in the time intervals of better performance. However, larger values of $K$ also enhance the impact of photon loss, necessitating a careful optimization of $K$ for fixed system parameters.

In Fig.~\ref{fig:fidelity}(a), we see that for $N_s^{\diamond}= 0.5$, the binomial code with $N=3$ for a loss rate of $\kappa/\nu=0.01$ performs reasonably well for all values of $r$ for Markovian to non-Markovian limits. In Fig.~\ref{fig:fidelity}(b), if we instead fix the value of $\tau$, we notice that $r=1$ yields a poor performance. This also corresponds to the noisiest state for a given $\tau$ in the inset of Fig.~\ref{fig:fidelity}(a), consequently leading to lower values of fidelity. Thus, the performance of these codes for a given $r$ depends on $N$ and $\tau$. 

In Fig.~\ref{fig:fidelity}(c), we consider infidelity vs. $n_{\text{code}}$ for $r = 0.001$ and zero loss, isolating the effect of dephasing at a fixed $\tau= 1$. Since canonical phase measurements, although ideal, are unphysical, a more practical approach for information about the phase distribution is provided by heterodyne detection for which the matrix elements of $B$ are not equal to one for $m \neq n$. An approximate scheme to implement matrix $B$ in Eq.\eqref{eq:povm} is the adaptive homodyne measurement (AHM), which relies on continuously adjusting the local oscillator phase $\theta$ based on measurement history~\cite{measurement2}, to reduce the information gain about photon distribution. We evaluate the performance for AHM and as shown in Fig.~\ref{fig:fidelity}, the performance with AHM closely matches the ideal canonical case to a very good extent.

We further compare these results with the performance of an optimal recovery map, obtained via a semidefinite program maximizing fidelity~\cite{arne, optimal}. 

An analytical upper bound on the average gate fidelity can be derived using a well-known inequality relating measurement outcomes to trace distance~\cite{Nielsen_Chuang_2010}:
\[
\frac{1}{2} \sum_m |p_m - q_m| \leq D(\hat{\rho}, \hat{\sigma}),
\]
where $p_m = \mathrm{Tr}(\hat{E}_m \hat{\rho})$ and $q_m = \mathrm{Tr}(\hat{E}_m \hat{\sigma})$ are POVM probabilities, and $D(\cdot,\cdot)$ denotes trace distance between the states.

Our case with $\hat{\rho} = \mathscr{N}(\ket{+}\bra{+})$ and $\hat{\sigma} = \mathscr{N}(\ket{-}\bra{-})$  yields the bound
\begin{equation}
    \mathcal{F}(\tau) \leq \frac{2 + [1 + D(\ket{+_\tau}_N, \ket{-_\tau}_N)] [1 + D(\ket{+}_M, \ket{-}_M)]}{6},
    \label{eq:fid_limit}
\end{equation}
where $\ket{\pm_\tau}_N$ are the evolved codewords at time $\tau$ after being subjected to the dephasing channel (we refer to Appendix~\ref{app:fidelity_derivation} for derivation). This equation shows that the fidelity is limited by the oscillatory behavior of the trace distance between the codewords, which also is key to evaluating the non-Markovian measure using BLP. The fidelity limit is plotted in Fig.~\ref{fig:fidelity}(c) where we observe that it closely follows the results of the optimal recovery map. 

To summarise, our results show that recovery fidelity under RTN dephasing depends sensitively on the order of symmetry $N$ and on the time at which the error correction is applied $\tau$. We observe oscillations in $ \mathcal{F}_{\text{EC}}$ due to non-Markovian effects, with optimal performance in highly non-Markovian ($r \leq 10^{-3}$) for certain timescales. We provide an understanding of these oscillations through an analytic expression derived for a purely dephasing channel. In the highly Markovian limit ($ r\geq 10^2$) continues to perform very well for a range of average photon number of RSB codes, reproducing the results of a Gaussian dephasing channel used in~\cite{arne,timo}.
For a given $\tau$, in the intermediate regime ($r \sim 1$), the average gate fidelity dips below the break-even threshold.
These findings are highly relevant, as experimental systems such as superconducting qubits and 3D cavities exhibit RTN parameters in the $r \sim 10^{-4}$ to $10^{-1}$ range~\cite{exptls1, exptls3d}.
For $1/f$ noise, a commonly encountered non-Gaussian source of dephasing, we find that the codes perform comparably to Gaussian models when the number of fluctuators is $N_f \gtrsim 10$.
However, the performance degrades for small $N_f$. 
This work is the first of its kind to study the impact of RTN dephasing on non-Gaussian states relevant to QEC.

\section{Discussion and conclusions}
\label{conclusions}

In this work, we investigated the RTN and $1/f$ noise on bosonic modes, where such noise arises from the interaction with one or more two-level systems (TLSs). This class of noise is especially relevant in contemporary quantum hardware platforms such as 3D cavities and superconducting resonators, where strong coupling between TLS and bosonic modes has been experimentally observed. The $1/f$ noise, prevalent across quantum devices, is modeled via either a single RTN fluctuator with distributed switching rates or a set of multiple independent fluctuators coupled to the system.

From an open quantum systems perspective, we characterized the non-Markovian nature of the induced dynamics using the BLP trace distance-based measure $\mathcal{N}_{\text{BLP}}$. We observed a crossover in dynamics from non-Markovian to Markovian behavior close to $r = 1$, where $r = \xi/\nu$ is the ratio of the RTN switching rate to the coupling strength of the fluctuator to the bosonic mode. In the non-Markovian regime ($r \ll 1$), CV states exhibit revivals in the evolution of trace distance due to information backflow. We showed numerically that, for Gaussian states, the non-Markovianity measure is maximized by specific pairs of coherent states with amplitude depending on the ratio $r$. Meanwhile, for non-Gaussian states, particularly those used in rotation-symmetric bosonic (RSB) codes, $\mathcal{N}_{\text{BLP}}$ can grow unboundedly with the code’s symmetry order $N$. In contrast, the Markovian regime ($r \gg 1$) and the effective Markovian limit induced by $1/f$ noise with many fluctuators ($N_f \gtrsim 10$) result in a monotonic degradation of trace distance with time, corresponding to the measure being equal to zero.

We then analyzed the performance of RSB codes—specifically binomial and cat codes—under RTN and $1/f$ noise in addition to a loss channel using a Knill-type teleportation-based QEC circuit. Our results show that encoding with $N > 1$ consistently outperforms trivial encoding ($N = 1$) in both Markovian and non-Markovian regimes.
However, in the non-Markovian limit, the average gate fidelity, which is a metric of performance, exhibits periodic oscillations in time, making QEC especially effective at specific time intervals. 
The frequency of these oscillations increases linearly with $N$, while their temporal width can be broadened by tuning the code parameters $K$ (binomial) and $\alpha$ (cat). 
These features allow fidelities to exceed the break-even threshold significantly, provided error correction is applied in synchrony with these peaks.

In the Markovian regime ($r \gg 1$), as well as under $1/f$ noise with a sufficient number of fluctuators ($N_f \gtrsim 10$), fidelity degrades monotonically with time. As a function of the average photon number of the code, we still obtain high fidelities above the break-even threshold, recovering the results of Gaussian dephasing \cite{timo}. We further validate the robustness of our findings by comparing the ideal canonical phase measurement with the more experimentally feasible adaptive homodyne detection, observing negligible differences in performance.

Lastly, we derived a semi-analytical bound on the QEC fidelity for a general purely dephasing channel and found that this bound is governed by the trace distance between codewords of the input encoding. In the non-Markovian limit, we are fundamentally limited by the oscillatory behavior in the fidelity. This result bridges our open-system analysis of non-Markovianity in bosonic systems with the results of quantum error correction, highlighting the interplay between noise structure, code symmetry, and error correction capability.

Our findings deepen the understanding of how  CV codes interact with RTN and $1/f$ noise and provide an understanding of the non-Markovianity induced by such noise models to help us enhance QEC performance in near-term quantum devices.

\section*{Author contribution}

A.U. performed most of the analytical and numerical calculations and wrote the largest part of the manuscript.
T.H. contributed significant insights that helped guide the research direction and provided valuable support in structuring and editing the manuscript.
R.G.A. performed analytical and numerical calculations. A.S. provided insights on open quantum systems and to the manuscript. G.F. conceived and supervised the project and wrote parts of the manuscript.

\section*{Acknowledgments}

We acknowledge useful discussions with Aditya Jayaraman, Axel Eriksson, Christophe Vuillot, Diptiman Sen, Kunal Helambe,  Nicolas Didier,  Simone Gasparinetti and Markus Hennrich.
We warmly thank Sabrina Maniscalco for helping in the stage of conception of this project.
G.F.\ acknowledges funding from the European Union’s Horizon Europe Framework Programme (EIC Pathfinder Challenge project Veriqub) under Grant Agreement No.\ 101114899.
G.F. acknowledges financial support from the Swedish Research Council through the project grant VR DAIQUIRI. G.F., T.H.  and A.U. acknowledge support from the Knut and Alice Wallenberg Foundation through the Wallenberg Center for Quantum Technology (WACQT). G.F., A.U. and R. G. A. acknowledge funding from Chalmers Area of Advance Nano.  A.S. acknowledges support from MUR under the PON Ricerca e Innovazione 2014–2020 project EEQU and under the PRIN 2022 Project ``Quantum Reservoir Computing (QuReCo)" (2022FEXLYB).

\appendix

\section{Derivation of dephasing function of RTN}
\label{app:dephasing_function}
The dephasing function provided in Eq.\eqref{eq:dephasingfunction} can be derived in two ways. We briefly derive it in both ways in this section. The average \( G(\tau) = \langle e^{i(m-n)\phi(\tau)} \rangle_{\phi(\tau)} \) can be computed by expanding the functional in time-ordered Taylor series, as~\cite{rtn_stat}:

\begin{equation}
G(\tau) = \sum_{k=0}^{\infty} (i (m-n))^k I_k(\tau),
\label{eq:G_expansion}
\end{equation}

where 

\begin{equation}
I_k(\tau) = \int_0^\tau d\tau_1 \int_0^{\tau_1} d\tau_2 \dots \int_0^{\tau_k} d\tau_k \langle c(\tau_1)c(\tau_2) \dots c(\tau_k) \rangle.
\end{equation}

Using the following relations from the statistics of the noise,  $\langle c(\tau_1) c(\tau_2) \rangle = e^{-2r |\tau_1 - \tau_2|}$ and  $\langle c(\tau_1) c(\tau_2) \dots c(\tau_n) \rangle = \langle c(\tau_1) c(\tau_2) \rangle \langle c(\tau_3) \dots c(\tau_n) \rangle$, we differentiate  Eq.\eqref{eq:G_expansion} twice to obtain a second-order differential equation for $G(\tau)$. The characteristic equation with $G(\tau)= e^{x\tau}$
gives $x^2 + 2r x + (m-n)^2 = 0$
 with the two solutions being $x_{\pm}= -r \pm \sqrt{r^2 - (m-n)^2} $ where $r$ is the ratio fo the switching rate of RTN $\xi$ to the coupling strength $\nu$. The initial conditions \( G(0) = 1 \) and \( G'(0) = 0 \) lead to the following solution:
\begin{equation}
G(\tau) = e^{-r\tau} \left( \cosh(\Omega \tau) + \frac{r}{\Omega} \sinh(\Omega \tau) \right).
\label{eq:derive_G}
\end{equation}

Alternatively, we can also derive the probability distribution for $\phi(\tau)$ by following the method used in~\cite{decohrtn2}. Let $p(\phi,\tau) = p_r(\phi,\tau)+p_l(\phi,\tau)$, where $p_r(\phi,\tau)$, $ p_l(\phi,\tau)$ are the probability densities that $c(\tau-\delta \tau)=\pm 1$ respectively, while $\phi(\tau):=\int_{0}^\tau\dd s\;c(s) = \phi$ and $\delta \tau$ is taken to be sufficiently small that only one flip can take place at most during $\delta \tau$. It is easy to see that
    \begin{align}
        p_r(\phi, \tau+\delta \tau) = (1-r\delta \tau) p_r(\phi - \delta \tau, \tau) + r\delta \tau\; p_l(\phi - \delta \tau, \tau)\nn\\
        p_l(\phi, \tau+\delta \tau) = (1-r\delta \tau) p_l(\phi + \delta \tau, \tau) + r\delta \tau \;p_r(\phi + \delta \tau, \tau). \nn
    \end{align}
    Expanding the functions $p_r$ and $p_l$ as a Taylor series of either of its arguments around $\phi$ and $\tau$ respectively, and taking the limit $\delta \tau\to 0$, we get two coupled differential equations for $p = p_r+p_l$ and $q:= p_r-p_l$. Eliminating $q$, one can find the \textit{telegraph equation}:
    \begin{align}
        p_{\tau\tau} + 2r p_\tau = p_{\phi\phi}.
    \end{align}
    A solution to this equation can be found, subject to the initial conditions $p(\phi,0) = \delta(\phi)$ and $p_\tau(\phi,0) = 0$ (where the second condition follows from the fact that $P(c(0)=\pm 1)=0.5$): $p(\phi,\tau) = \int \frac{\dd k}{2\pi}\l(a^+_ke^{-i\omega^+_k\tau}+a^-_ke^{-i\omega^-_k\tau}\r)e^{ik\phi}$, $\omega^\pm_k= -ir \pm \sqrt{k^2 - r^2}$, $a^\pm_k = \frac{1}{2}\l(1\pm\frac{ir}{\sqrt{k^2-r^2}}\r)$. 

    The dephasing factor due to RTN can be directly calculated as follows
    \begin{align}
        G(m,n,\tau) &= \int_{-\infty}^{\infty}p(\phi,\tau)e^{i(m-n)\phi}\nn\\
         &= \l. a^+_ke^{-i\omega^+_k\tau}+a^-_ke^{-i\omega^-_k\tau}\r\vert_{k=n-m}\nn\\
         &= e^{-r\tau} \left( \cosh(\Omega \tau) + \frac{r}{\Omega} \sinh(\Omega \tau) \right),
    \end{align}
in agreement with Eq.\eqref{eq:derive_G}.

\section{Derivation of two limits of evolved density matrix under a single RTN fluctuator}
\label{app:two_limits}
In this Appendix, we derive the approximate expressions for the density matrix under RTN dephasing in the two limits of the value of $r$: $r\ll 1$ and $r\gg 1$. For $r\ll 1$, i.e, where the average switching rate of the fluctuator is smaller than the coupling strength, we see in Sec.~\ref{sec:rtn_intro} that the evolution of a coherent state has the majority of the contribution coming from two blobs going around the circle for small times. We can understand this behaviour by expanding the expression of the dephasing function in order of $r$. From Eq.\eqref{eq:dephasingfunction} we have, 

\begin{eqnarray}
    \langle e^{i\phi(m-n)} \rangle_{\phi(\tau)} &=& e^{-r\tau} \left ( \cos~ |\Omega|\tau + \frac{r}{|\Omega|}\sin ~|\Omega| \tau \right)  \nonumber \\
    &=& e^{-r\tau} \left ( (\frac{|\Omega| + r}{2|\Omega|})e^{i|\Omega| \tau} + (\frac{|\Omega| - r}{2|\Omega|})e^{-i|\Omega| \tau})\right), \nonumber \\ 
\end{eqnarray}
where $\Omega = \sqrt{(m-n)^2 - r^2 }$.

Plugging this into the expression of the evolved density matrix and keeping the terms only up to zeroth order in $r$, we have 
\begin{align}
     \hat{\rho}(\tau) &=&  e^{-r \tau}  (\frac{|\Omega| + r}{2|\Omega|}) e^{-|\alpha|^2}\sum_{m,n=0}^{\infty} \frac{(\alpha)^n (\alpha^{*})^m e^{i|\Omega|\tau}}{\sqrt{n! m!}}\ket{n} \bra{m}  \nonumber \\
     &+&  e^{-r \tau} (\frac{|\Omega| - r}{2|\Omega|}) e^{-|\alpha|^2}\sum_{m,n=0}^{\infty} \frac{(\alpha)^n (\alpha^{*})^m e^{-i|\Omega| \tau}}{\sqrt{n! m!}}\ket{n} \bra{m}, 
\end{align}
   
Further for small times when $r\tau \ll 1$, see that $e^{-r \tau} \approx 1$, and the expression becomes 
\begin{eqnarray}
     \hat{\rho}(\tau) &=& \frac{1}{2}  e^{-|\alpha|^2}\sum_{m,n=0}^{\infty} \frac{\alpha_0^n \alpha_0^m e^{i(m-n)(\theta+\tau)}}{\sqrt{n!m!}}\ket{n} \bra{m} \nonumber \\
     &+& \frac{1}{2}  e^{-|\alpha|^2}\sum_{m,n=0}^{\infty} \frac{\alpha_0^n \alpha_0^m e^{i(m-n)(\theta-\tau)}}{\sqrt{n! m!}}\ket{n} \bra{m},
\label{Eq:RTN_two_blob}
\end{eqnarray}   
where $\alpha = \alpha_0 e^{i\theta}$.
Now we observe that as a function of $\tau$, the contribution to the evolved state equally comes from two coherent states with their phases shifted by $\tau$. Thus, the Wigner function has two blobs that move around the circle as a function of time. As $\tau$ increases, the term $e^{-r\tau} $ can no longer be approximated to be 1, and it begins to give a decaying component to the density matrix. After a long time, the effect of this RTN noise is similar to that of Gaussian noise, where the state is completely dephased.

In the other limit, where we have $r \gg 1, m,n$, we see that the evolved density matrix tends to the case of a Gaussian dephasing of the form $\mathcal{N}_{\sigma^2} \hat{\rho}_0 = \int_{-\infty}^{\infty} d\theta~ p(\theta) e^{-i\theta \hat{n}} \hat{\rho}_0 e^{i\theta \hat{n}}$ where $p(\theta)$ is taken from a Gaussian distribution with standard deviation $\sigma^2= k_{\phi}\tau$. For the Gaussian channel, by expanding the density matrix in the Fock basis, and then performing the integral over $\theta$ in the above equation, and after completing the squares, we get
\begin{eqnarray}
    \mathcal{N}_{\sigma^2} \hat{\rho}_0 &=&  \sum_{m,n=0}^{\infty} \int_{-\infty}^{\infty} \hspace{-1em} d\theta \frac{1}{\sqrt{2\pi \sigma^2}}e^{-\frac{\theta^2}{2\sigma^2}} e^{i\theta (m-n)} \rho_{m,n}\ket{m}  \bra{n} \nonumber \\
    &=&\sum_{m,n=0}^{\infty} \rho_{m,n}\ket{m} \bra{n}  e^{-\frac{1}{2}(m-n)^2\sigma^2},
    \label{Eq:gau_rho}
\end{eqnarray}
where $\rho_{m,n}= \bra{m}\hat{\rho}_0\ket{n}$.
We now show how the evolution of the density matrix under RTN with a high switching rate reduces to this result with $\sigma^2 =  \tau / r$. In this limit of $r \gg 1, m-n$, we have $\Omega= r(1-\frac{(m-n)^2}{2r^2})$ and $\cos h\Omega \tau = \sin h\Omega \tau \simeq e^{\Omega \tau }/2$. Up to zeroth order in $\frac{(m-n)^2}{2r^2}$, Eq.\eqref{eq:rho_evolved} becomes

\begin{eqnarray}
      \hat{\rho}(t) &=& \sum_{m,n= 0}^{\infty}  e^{-r \tau} e^{\Omega \tau}  \rho_{m,n} \ket{m} \bra{n} \nonumber \\
        &=& \sum_{m,n= 0}^{\infty}  e^{- \frac{1}{2}\frac{(m-n)^2 \tau}{r}}  \rho_{m,n} \ket{m} \bra{n}.
\end{eqnarray}

Comparing this with Eq.\eqref{Eq:gau_rho}, we see that it has a form similar to the evolved Gaussian dephased state with $k_{\phi} = 1/r$.

\section{Derivation of the density matrix under the influence of multiple fluctuators}
\label{app:multiple_fluctuator_derivation}
 We have obtained the analytical expression for the decoherence due to $1/f$ noise in a bosonic system in Eq.\eqref{eq:int_multiple}. The following integral appears,
\begin{align}
    \int_{r_{min}}^{r_{max}}\frac{\dd r}{r}e^{-r \tau}\l( \cosh \Omega \tau + \frac{r}{\Omega}\sinh \Omega \tau \r),
\end{align}
where $\Omega = \sqrt{r^2-a^2}$ for $a = (m-n) $. Writing $r_{min}= r_m$ and $r_{max}= r_M$ for convenience, and assuming $r_m <a<r_M$, we split this integral into two parts, with $I= I_1 + I_2$.
\begin{align}
    I_1 \equiv \int_{r_m}^{a}\frac{\dd r}{r}e^{-r\tau}\l( \cos \Omega^{\prime} \tau + \frac{r}{\Omega^{\prime} }\sin \Omega^{\prime} \tau\r);\\
    I_2 \equiv \int_{a}^{r_M}\frac{\dd r}{r}e^{-r\tau}\l( \cosh \Omega \tau + \frac{r}{\Omega}\sinh \Omega \tau\r),
\end{align}
where in $I_1$, $\Omega^{\prime}  = \sqrt{a^2-r^2}$. 
Performing the integral for the three cases, we obtain \\
$
I=\begin{cases}
			F(z_M^+)+F(z_M^-) - F(z_m)-F(z_m^*), & r_m<a<r_M \\
            F(z_M^+)+F(z_M^-) - F(z_m^+)-F(z_m^-), & a<r_m<r_M \\
            F(z_M)+F(z_M^*) - F(z_m)-F(z_m^*) & r_m<r_M<a,
		 \end{cases}
$
\\
where $F(x)= E_1(x) - \frac{e^{ia}}{2}E_1(x+ia)-\frac{e^{-ia}}{2}E_1(x-ia)$, $z_{M/m}= r_{M/m} -i\sqrt{a^2-r_{M/m}^2}$, $  z^{\pm}_{M,m} = r_{M/m} \pm \sqrt{r_{M/m}^2-a^2}$ and 
 $E_1(x)$ is called the exponential integral defined in~\cite{abramowitz} and is readily available in the \href{https://docs.scipy.org/doc/scipy/reference/special.html}{scipy.special} python package. To get the final decoherence function at time $\tau$, we divide the results above by $\ln\l(r_M/r_m\r)$ and raise them to the power of $N_f$. Additionally, when $a = 0$ or $\tau = 0$, some of these terms will diverge; however, as we know that in either of these cases, no dephasing occurs even for a single fluctuator. The decoherence function can safely be set to $1$ in either of the cases.

\section{Optimisation of non-Markovian measure for Gaussian states}
\label{app:coherent_state_maximisation}

\begin{figure*}
    \centering
    \includegraphics[width=\linewidth]{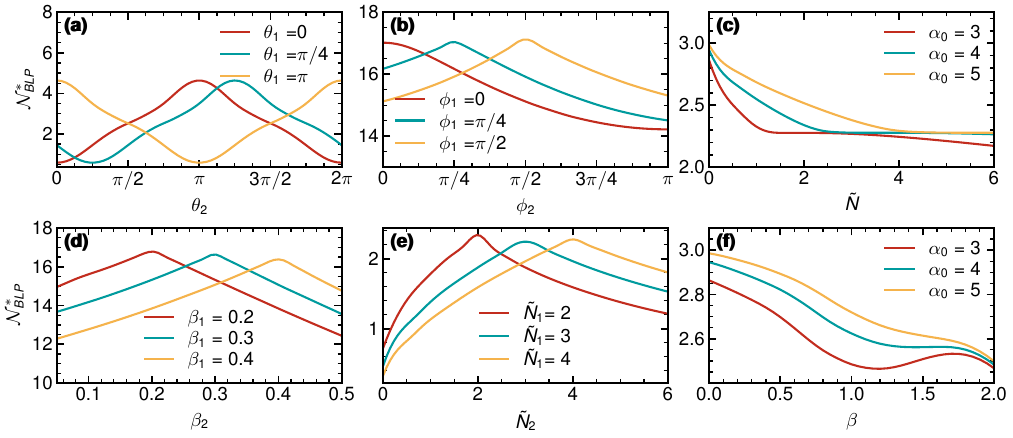}
    \caption{The optimisation procedure consists of maximising over the following ten parameters: $(\alpha_{0_1}, \alpha_{0_2}, \theta_1, \theta_2, \beta_{0_1}, \beta_{0_2}, \gamma_1, \gamma_2, \tilde{N}_1, \tilde{N}_2)$. We see that for different values of the maxima in the plots occur at (a) $|\theta_1 - \theta_2| = \pi$, (b) $\phi_1= \phi_2$, (d) $\beta_1= \beta_2$ and (e) $\tilde{N}_1 = \tilde{N}_2$. Further in (c) and (f), we see that squeezing and thermal excitations do not increase the value of $\mathcal{N}_{\text{BLP}}$. }
\label{fig:optimisation}
\end{figure*}

 In this Appendix, we elaborate on the optimization procedure of the non-Markovian measure for Gaussian states. As discussed in Sec.~\ref{sec:RTN-Gaussian}, a general Gaussian state is described by two complex parameters $\alpha$ and $\beta$ and one real parameter $\tilde{N}$ (See Eq.\eqref{eq:general_gaussian}). To find the value of $\mathcal{N}_{\text{BLP}}$ for the subset of Gaussian states, we need to maximize over two initial states and thus over the following ten parameters: $(\alpha_{0_1}, \alpha_{0_2}, \theta_1, \theta_2, \beta_{0_1}, \beta_{0_2}, \gamma_1, \gamma_2, \tilde{N}_1, \tilde{N}_2)$.
 We observe that for certain relations between these parameters, the non-Markovian measure is maximum and gives the same value for a set of these parameters. By fixing the parameters of the first state, we see that the parameters of the second state satisfying the following conditions give the maximum value of $\mathcal{N}_{\text{BLP}}^{*}$ (Fig.~\ref{fig:optimisation}(a), (b), (d), (e)), 
\begin{eqnarray}
\label{Eq:conditions}
    \alpha_{0_2}&=& \alpha_{0_1} ,\nonumber \\
    \theta_2 &=& \pm \pi - \theta_1, \nonumber \\
    \beta_{0_2} &=& \beta_{0_1}, \nonumber \\
    \gamma_2 &=& \gamma_1, \nonumber \\
    \tilde{N}_2 &=& \tilde{N}_1.
\end{eqnarray}
The value of the measure is the same for any choice of the parameters $\theta_2$ and $\gamma_2$ that satisfy the above conditions. 
Thus the optimisation problem reduces to considering three parameters $\alpha_{1_0}= \alpha_{2_0}= \alpha_0$, $\beta_{1_0}= \beta_{2_0}= \beta$ and $\tilde{N}_1 = \tilde{N}_2 = \tilde{N}$. Further, as a function of the squeezing amplitude and the average photon number, we observe that the maximum value of $\mathcal{N}_{\text{BLP}}^{*}$ scanning over all states is obtained for a pair of coherent states with a particular value of $\alpha_0$ depending on the ratio $r$ with no squeezing and no thermal fluctuations respectively (shown in Fig.~\ref{fig:optimisation}(f) and (c)) and Fig.~\ref{fig:main_results2}(a)).

\section{ $\mathcal{N}_{\text{BLP}}$ measure for non-Gaussian states} 
\label{app:nblp_non_gaussian}
 In this Appendix, we show that going beyond the Gaussian states, one can find a limiting case, where a pair of non-Gaussian states will yield an integrated revival of trace distance, which is unbounded from above. To see this, let us take the subspace $H_l$ spanned by the Fock states  $\{|0>, |l>\}$ of the infinite dimensional Hilbert space, comprising of all normalised linear combinations of the Fock states $\{\ket{0},\ket{l}\}$. In the non-Markovian regime of RTN, any state belonging to the space $\mathcal{S}(H_l)$ will undergo decoherence exactly like a qubit but with a dephasing factor given by 
\begin{align}
    G(l,\tau) = e^{-r \tau}\l(\cosh \Omega_l \tau + \frac{r}{\Omega_l} \sinh \Omega_l \tau \r),
\end{align}
where $\Omega_l = \sqrt{r^2-l^2}$. As it has been shown in~\cite{opt.tr.dist}, the trace distance for the optimal pair of qubit states (which maximizes $\mathcal{N}_{\text{BLP}}$) evolves as $D(\tau) = \abs{G(l,\tau)}$. The revival of the trace distance, quantified as the cumulative sum of the trace distance between two specific states for all the time intervals during which the trace distance is increasing can be calculated and is given by 
\begin{align}
    N_l = \sum_{n=1}^\infty \abs{G(l,T^n_l)},
\end{align}
where $T^n_l = n\pi/\Omega_l$ is the time when the $n$-th maximum of $\abs{G(l,\tau)}$ occurs (the minima are all zero). This geometric series yields
\begin{align}
    N_l = \frac{1}{e^{\pi r/\Omega_l}-1}.
\end{align}
For every $r$, in the limit $l\to \infty$, $N_l \to \infty$. Hence, the BLP measure for RTN affecting a bosonic system is unbounded.

\section{Wigner negativity - another measure of non-Markovianity}
\label{app:wigner_neg}

Another measure of non-Markovianity based on the Wigner negativity volume also serves as an indicator of the quantum correlations in any composite quantum system \cite{wigneg1, wigneg2, wigneg3,wigneg5, wigneg6}.  In this Appendix, we use Wigner negativity as an alternative metric to further validate our findings regarding the transition between Markovian and non-Markovian regimes for the RTN noise, characterized by the ratio $r$. 
The generalised Wigner function negativity volume is defined as \cite{wigneg3}
\begin{equation}
    N_V = \frac{1}{2} \l[ \int dq dp \; |W_{\hat{\rho}}(q,p)| -1 \r],
    \label{eq:wigner_neg}
\end{equation}
where  $W_{\hat{\rho}}(q,p)$ is the Wigner function defined over the phase space as
\begin{align}
    W_{\hat{\rho}}(q,p) = \frac{1}{\pi}\int_{-\infty}^\infty\dd y\; e^{2ipy} \bra{q+y}\hat{\rho}\ket{q-y},
\end{align}
where $\ket{q}$ is an eigenstate of the position operator, $\hat{q}$: $\hat{q}\ket{q} = q\ket{q}$.

For Markovian dynamics, the negativity volume is a monotonically decreasing function of time, reflecting the increasing entanglement between the system and the environment. In contrast, non-Markovian dynamics are characterized by an increase in Wigner negativity at certain time intervals. This oscillatory behavior enables the quantification of non-Markovianity based on the Wigner negativity volume.
The degree of non-Markovianity $\mathcal{N}_{WN}$ is defined as \cite{wigneg4}
\begin{equation}
    \mathcal{N}_{WN} = 1 -\frac{|\int dt \ \frac{dN_V}{dt}|}{\int dt \ |\frac{dN_V}{dt}|}.
    \label{eq:nwn}
\end{equation}
By definition, this measure is zero again for Markovian processes as  $ \frac{dN_V}{dt}<0 $  for all times. 

\begin{figure*}
    \centering
    \includegraphics[width=\linewidth]{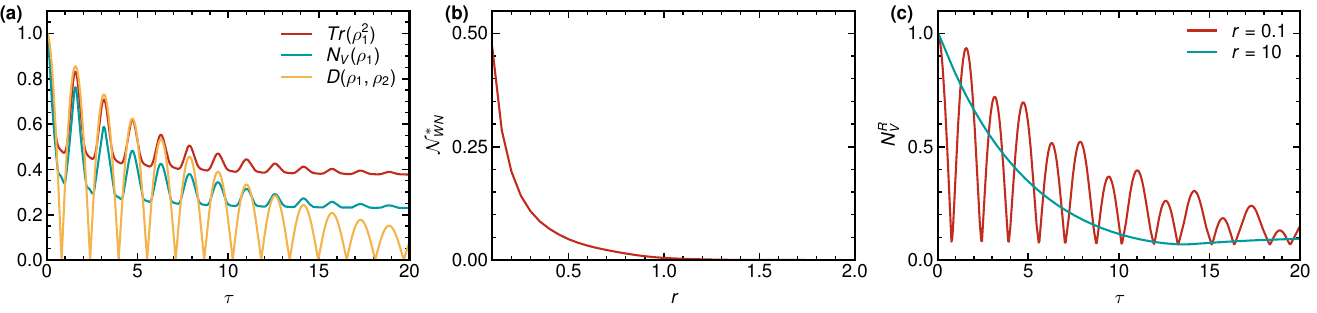}
     \caption{(a) The Wigner negativity exhibits oscillations with $\tau$ for a binomial state $\ket{+}_{2,2}$ in the non-Markovian regime where $r=0.1$. The period of oscillations of Wigner negativity is the same as that of the purity of $\ket{+}_{2,2}$ and the trace distance between states $\ket{+}_{2,2}$ and $\ket{-}_{2,2}$. (b) We plot the measure of non-Markovianity based on Wigner negativity $N_{WN}^{*} $ for the same state $\ket{+}_{2,2}$ as a function of the ratio of $r$. The asterisk in the superscript indicates that this measure is not maximised over a set of states, but is specific to a given state. From this measure, we see the bosonic mode transits from a non-Markovian to a Markovian regime close to $r=1$. (c) The plot shows Wigner negativity integrated only along a ring at $R= 1.2$ for $\ket{+}_{1,2}$ in the phase space, and we again observe oscillations in the non-Markovian regime and a monotonic decay in the Markovian regime.  }
    \label{fig:wigner_neg}
\end{figure*}

From an experimental perspective, studying Wigner negativity offers significant advantages. Its evolution is easier to probe and measure compared to the $\mathcal{N}_{\text{BLP}}$ measure of non-Markovianity, making it a practical choice for experimental validation of non-Markovian effects. 
In Fig.~\ref{fig:wigner_neg} (a) we plot the evolution of the Wigner negativity as given in Eq.(\ref{eq:wigner_neg}) for a binomial state $\ket{+}_{2,2}$. In the non-Markovian regime when $r=0.1$, we see revivals of $N_V$, while for $r=10$, the function monotonically decays. The period of oscillations of $N_V$ is the same as the period of revivals of the trace distance for the same state with $\ket{-}_{2,2}$. However, we notice that the Wigner negativity tends to a constant value faster than the trace distance approaches zero. We plot this in Fig.~\ref{fig:wigner_neg} (b) and as expected the measure exponentially reduces as a function of $r$ approaching the Markovian limit. 
Further, to facilitate the study of the behaviour of the Wigner negativity experimentally, we also look at the Wigner function integrated over a ring of radius $\sqrt{p^2 + q^2} = R$ in the phase space. We define this quantity $N_V (R)$ as follows: 

\begin{equation}
    N_V^R = \frac{1}{2} \l[ \int d\theta \; |W_{\hat{\rho}}(R,\theta)| -1 \r],
\end{equation}
where $R= \sqrt{p^2 + q^2}$ and $\theta = \tan^{-1}\frac{p}{q}$. 
In Fig.~\ref{fig:wigner_neg} (c), we plot how Wigner negativity along a ring of radius $R= 1.2$ for a binomial state $\ket{+}_{1,2}$ as a function of time. The difference in behaviour in the non-Markovian and the Markovian limits is evident even when integrated only along a ring; the radius of the ring has to be chosen in accordance with the given initial state, such that the Wigner negativity is finite. 

\section{Derivation of the recovery map of Knill error correction circuit}
\label{app:circuit_derivation}
In this Appendix, we derive the result of the recovery channel of the Knill error correction circuit given in~\cite{arne}. Referring back to the circuit in Fig.~\ref{fig:circuit_ec}, the input state of the three modes has the following form, 

\begin{equation}
    \mathscr{N}_N \otimes \mathds{1}_{ML} (\hat{\rho}_N \otimes \ketbra{+}{+}_M \otimes \ketbra{+}{+}_L),
\end{equation}
where $\mathds{1}_{ab}$ is the identity operator acting on the modes $a$ and $b$, and $\mathscr{N}_N$ is the map of the noise channel acting on order-$N$ input mode. We henceforth omit the identity operator, and it is understood by context on the modes it acts on. 
Here, the input state in the computational code basis is given by 
$\hat{\rho}_N = \sum_{p,q=0}^{1} {\rho}_{p,q}\ket{p}_N \bra{q}_N$.  
The state, after passing through the two controlled rotation gates, is

\begin{equation}
      \hat{C}_{ML} \hat{C}_{NM} \mathscr{N}_N (\hat{\rho}_N \ketbra{+}{+}_M \ketbra{+}{+}_L) \hat{C}_{NM}^{\dagger} \hat{C}_{ML}^{\dagger},
\end{equation}
    where $\hat{C}_{NM}$ denotes the CROT gate acting on modes of order $N$ and order $M$.  Inserting $\hat{C}_{NM} \hat{C}_{NM}^{\dagger}$ we can rewrite the above equation as follows, 
\begin{tiny}
\begin{equation}
        \hat{C}_{ML} \overbrace{\hat{C}_{NM} \mathscr{N}_N  \hat{C}_{NM}^{\dagger} \underbrace{\hat{C}_{NM} \hat{\rho}_N \ketbra{+}{+}_M \ketbra{+}{+}_L \hat{C}_{NM}^{\dagger}}_{\hat{A}(\hat{\rho})} \hat{C}_{NM} \hat{C}_{NM}^{\dagger}}^{\hat{B}(\hat{\rho})} \hat{C}_{ML}^{\dagger},  \\ 
\end{equation}
\label{rhoout}
\end{tiny}
where we have used $\hat{\rho}_N = \sum_{p,q=0}^{1} {\rho}_{p,q}\ket{p}_N \bra{q}_N$.

 By looking at the action of $\hat{C}_{NM}$ on the input states, we can evaluate the term $\hat{A}(\hat{\rho})$. 
We have 
\begin{equation}
    \hat{C}_{NM} \ket{p}_N \ket{+}_M = \ket{p}_N \ket{h^p}_M,
\end{equation}
where $\ket{h^0}_M= \ket{+}_M$ and $\ket{h^1}_M= \ket{-}_M$ for $p=0$ and $p=1$ respectively. Thus we have 
\begin{equation}
    \hat{A}(\hat{\rho}) =  \sum_{p,q=0}^{1} {\rho}_{p,q}\ketbra{p}{q}_N \ketbra{h^p}{h^q}_M \ketbra{+}{+}_L.
\end{equation}
Next we look at the term $\hat{B}(\hat{\rho})$ and implementing the action of the noise channel in terms of the Kraus operators for both loss and dephasing $\hat{K}_l$ and $\hat{K}_d$ respectively, we have 
\begin{eqnarray}
     \hat{B}(\hat{\rho}) &=& \hat{C}_{NM} \mathscr{N} (\hat{C}_{NM}^{\dagger} \hat{A}(\hat{\rho}) \hat{C}_{NM}) \hat{C}_{NM}^{\dagger} \nonumber \\    
     &=& \sum_{l,d} \hat{K_d} \hat{K_l} \hat{R}_M  \hat{A}(\hat{\rho}) \hat{R}_M^{\dagger} \hat{K_l}^{\dagger} \hat{K_d}^{\dagger},
\end{eqnarray}
where we have used the fact that CROT gate commutes with the dephasing Kraus operator and gives an additional rotational operation $\hat{R}_M$ on the second mode when it is commuted through the loss Kraus operator~\cite{arne}. The angle of this rotation operator depends on the index $l$ of the loss Kraus operator~\cite{arne}. The final equation thus gives, 
\begin{equation}
    \hat{C}_{ML} \hat{B}(\hat{\rho}) \hat{C}_{ML}^{\dagger} =  \sum_{l,d} \hat{K_d} \hat{K_l} \hat{R}_M  \hat{C}_{ML} \hat{A}(\hat{\rho}) \hat{C}_{ML}^{\dagger} \hat{R}_M^{\dagger} \hat{K_l}^{\dagger} \hat{K_d}^{\dagger},
    \label{eq:intermediate}
\end{equation}
where we have used the fact that the operator $\hat{C}_{ML}$ does not act on the order-$N$ mode, it commutes through the Kraus and the error operators.

We now look at the action of $\hat{C}_{ML}$ on the state for $p=0$ and $p=1$ and using Pauli operators acting on the order-$L$ state we see that
\begin{eqnarray}
    \hat{C}_{ML} \ket{0,+,+} &=& \frac{1}{2} ( \ket{+,+}\mathds{1}+\ket{-,+}\hat{Z} \nonumber \\
    &+&\ket{+,-}\hat{X} + \ket{-,-}\hat{X}\hat{Z}) \ket{0}_L \nonumber \\
    \hat{C}_{ML} \ket{1,-,+} &=&  \frac{1}{2} ( \ket{+,+}\mathds{1}+\ket{-,+}\hat{Z} \nonumber \\ 
    &+& \ket{+,-}\hat{X} + \ket{-,-}\hat{X}\hat{Z}) \ket{1}_L,
\end{eqnarray}
where $\ket{a,b,c}= \ket{a}_N \ket{b}_M \ket{c}_L$ with $a,b,c= 0, 1, \pm$. 

Using the following notation, $\ket{0}= \ket{+,+}, \ket{1}= \ket{+,-}, \ket{2}=\ket{-,+} $ and $\ket{3}= \ket{-,-}$ for the states, and $\hat{P}_i \in \{ \mathds{I},\hat{X},\hat{Z}, \hat{X}\hat{Z}\}$, we get 
\begin{equation}
    \hat{C}_{ML} \hat{A}(\hat{\rho}) \hat{C}_{ML}^{\dagger} = \frac{1}{4} \sum_{i,j=0}^{3} \ket{i} \bra{j} \hat{P}_i \hat{\rho}_L \hat{P}_j^{\dagger}.
\end{equation}
Thus finally Eq.~\eqref{eq:intermediate} after measurement of $\hat{M}_{x_1}$ and $\hat{M}_{x_2}$, and summing over all the values of $x_1$ and $x_2$, becomes
\begin{equation}
\hat{\rho}_{out} = \frac{1}{4} \sum_{x_1, x_2} \sum_{i,j =0}^{3} c_{ij}(\Vec{x})\hat{P}_i \hat{\rho}_L \hat{P}_j^{\dagger},
\end{equation}
where $c_{ij} = \Tr(\hat{M}_{x_1} \otimes \hat{M}_{x_2}(\hat{\mathcal{U}}_c \circ \mathscr{N} \circ \hat{\mathcal{U}}_c^{\dagger} \ket{i}\bra{j})) $ and $\hat{\mathcal{U}}_c \circ = \hat{C}_{NM} \circ \hat{C}_{NM}^{\dagger}$.

\section{Derivation of average gate fidelity for a purely dephasing channel}
\label{app:fidelity_derivation}
Here we include the derivation of the average gate fidelity of the recovery map in the Knill EC in the presence of a purely dephasing map and an arbitrary rotation-symmetric bosonic encoding. The $c_{ij}(x_1,x_2)$ matrices, given in the previous section, for a canonical phase measurement (where we have set the elements of the matrix $B$ in Eq.\eqref{eq:povm} to 1) can be explicitly written as
\begin{align}\label{can_c_mat}
     &c_{ij}(\phi_1,\phi_2)\nn\\
     = &\frac{1}{2\pi}\sum_{k,l=0}^\infty(-1)^{ka+lc}f_{kN}\;f^*_{lN} \;G(kN,lN,\tau) \;e^{i(l-k)N\phi_1}
    \nn\\
    &\cdot\frac{1}{2\pi}\sum_{k,l=0}^\infty(-1)^{kb+ld}g_{kM}\;g^*_{lM}\;e^{i(l-k)M\phi_2},
\end{align}
where $ ab$ and $cd$ are the binary representations of $i$ and $j$ respectively. The $f$ and $g$ are the binomial coefficients of the codewords, and the phases $\phi_1$ and $\phi_2$ are the measurements on the order-$N$ and order-$M$ code, respectively. $G(kN,lN,r,\tau)$ is the dephasing function with the $kN$ and $lN$ being the indexes of the Fock basis running over the integers. 
The average gate fidelity for a qubit system ($d=2$) is given by~\cite{av_gate_fid}
\begin{small}
\begin{align}
    \mathcal{F} = \frac{\Tr[\mathcal{E}(\mathbb{I})]+\Tr[\hat{X}\mathcal{E}(\hat{X})]+\Tr[\hat{Z}\mathcal{E}(\hat{Z})]+\Tr[\hat{Z}\hat{X}\mathcal{E}(\hat{X}\hat{Z})]+4}{12},
    \label{eq:av_gate_fidelity}
\end{align}
\end{small} where $\mathcal{E}$ is the map given in Eq. \eqref{eq:recovery_map}. Replacing the sum over $x_i$'s with integrals over $\phi_i$'s in this equation, we evaluate the contribution from each of these terms to the fidelity.  The first term $\Tr[\mathcal{E}(\mathbb{I})]$ trivially simplifies to 2.  
The second term can be computed as follows:
\begin{align}
    \Tr[\hat{X}\mathcal{E}(\hat{X})] \nn\\
    = &\frac{1}{4}\int\dd \phi_1\dd\phi_2 (c_{00}(\vec{\phi})+c_{11}(\vec{\phi})-c_{22}(\vec{\phi})-c_{33}(\vec{\phi}))\nn\\&\cdot \Tr\l[\hat{X}\hat{P}^{L\d}_{i*(\vec{\phi})}\hat{X}\hat{P}^L_{i*(\vec{\phi})}\r].
    \label{fid_X}
\end{align}
Writing $i$ as the binary equivalent $i \equiv ab$, we have
\begin{equation*}
c_{ii}(\vec{\phi})= c^{\tau}_a(\phi_1)\cdot c_b(\phi_2),
\end{equation*}
where $c^{\tau}_a(\phi_1)=\Tr\{\hat{M}(\phi_1)\mathscr{N}(\ketbra{(-)^a}{(-)^a}_N)\}$ and $c_b(\phi_2)= \Tr\{\hat{M}(\phi_2)(\ketbra{(-)^b}{(-)^b}_M))\}$. This allows us to rewrite 
\begin{equation}
    (c_{00}+c_{11}-c_{22}-c_{33})(\vec{\phi}) = (c^{\tau}_0-c^{\tau}_1)(\phi_1)(c_0+c_1)(\phi_2).
\end{equation}

\begin{figure}
    \centering
    \includegraphics[width=0.8\linewidth]{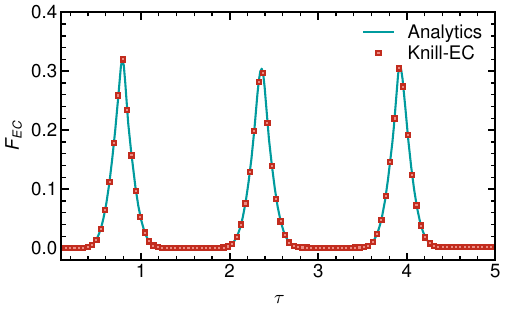}
    \caption{The analytical expression for the average gate fidelity of the circuit compared with the same obtained from the numerical realization of the error correction circuit}
    \label{fig:Fidelity_analytics}
\end{figure}

When $i^*(\vec{\phi})\in \{0,1\}$, the trace in Eq. \eqref{fid_X} evaluates to $2$ and $c^{\tau}_0(\phi_1)> c^{\tau}_1(\phi_1)$, otherwise the trace evaluates to $-2$ and $c^{\tau}_1(\phi_1)> c^{\tau}_0(\phi_1)$. Putting these together and performing the integral over $\phi_2$, which evaluates to 1, the expression simplifies to 

\begin{align}
    \Tr[\hat{X}\mathcal{E}(\hat{X})] = \int_0^{2\pi}\dd \phi_1\; \abs{c^{\tau}_0-c^{\tau}_1}(\phi_1).
\end{align}
A similar analysis shows that
\begin{align}
    \Tr[\hat{Z}\mathcal{E}(\hat{Z})] = \int_0^{2\pi}\dd \phi_2\; \abs{c_0-c_1}(\phi_2),
\end{align}
where we have assumed the noise is trace-preserving while evaluating the integral over $\phi_1$. We also have, 
\begin{align}
    \Tr[\hat{Z}\hat{X}\mathcal{E}(\hat{X}\hat{Z})] = &\frac{1}{2}\int_0^{2\pi}\dd \phi_1\; \abs{c^{\tau}_0-c^{\tau}_1}(\phi_1) \nonumber \\ 
    &\cdot \int_0^{2\pi}\dd \phi_2\; \abs{c_0-c_1}(\phi_2).
\end{align}
We substitute these into Eq.\eqref{eq:av_gate_fidelity} to get the final expression for fidelity of the error-corrected map. In Fig.~\ref{fig:Fidelity_analytics}, we compare the results from the values of $\mathcal{F}_{\text{EC}}$ obtained from the error circuit and the values from the integrals of the expression above, and find that they agree well. We note that for the final expressions, we can conveniently generalize them for any measurement scheme by keeping the elements of matrix $B$ throughout the proof. 
To infer the time dependence of the fidelity, it is sufficient to analyze $\int_0^{2\pi}\dd \phi_1\; \abs{c^{\tau}_0-c^{\tau}_1}(\phi_1)$ for a canonical measurement. For a binomial code with $K=2$ and arbitrary $N$, we see that the quantity
\begin{align}
    &\int_0^{2\pi}\dd \phi_1\; \abs{c^{\tau}_0-c^{\tau}_1}(\phi_1)\nn\\ = &\frac{4}{2\pi}\int_0^{2\pi}\dd \phi_1\; \abs{(f_0f_N+f_Nf_{2N})G(0, N,\tau)\cos(N\phi_1)}\nn\\ = &\frac{4\sqrt{2}}{\pi}\abs{G(0, N, \tau)},
    \label{eq:K_2_dephasing}
\end{align}
 oscillates with the frequency $\Omega = 2\sqrt{N^2-r^2}$. This matches the numerical simulation as well, where we observe the frequency of oscillations increase linearly with $N$. To illustrate what happens for higher $K$, let us take the example of $K = 3$, for which the integral becomes
\begin{align}
    &\int_0^{2\pi}\dd \phi_1\; \abs{c^{\tau}_0-c^{\tau}_1}(\phi_1)\nn\\ = &\frac{4}{2\pi}\int_0^{2\pi}\dd \phi\; \abs{F_1G(0, N,\tau)\cos(\phi) + F_3G(3N,t)\cos(3\phi)},
\end{align}
where $F_1 = (1/8)(2\sqrt{3}+3), F_3 = (1/8)$. We see that we have a higher contribution from $G(N,t)$ than from $G(3N,t)$, dominating the time dependence of fidelity, however, with a higher width of the maxima and a lesser value of minima. With higher values of $K$, this trend seems to be continuing as we checked numerically (see Fig.~\ref{fig:Fidelity_analytics}). However, when we include the loss channel, we encounter a trade-off since, with increasing $K$, the average excitation number also increases, leading to a higher photon loss rate.

\begin{figure*}
    \centering
    \includegraphics[width=\linewidth]{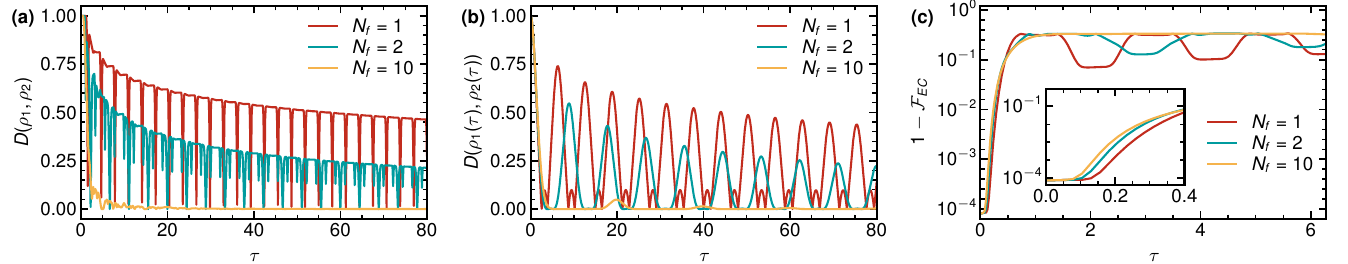}
    \caption{
 Evolution of the trace distance between pairs of quantum states: (a) coherent states with $\alpha_0 = 2$ and a phase difference of $\pi$, and (b) binomial states $\ket{+}_{1,2}$ and $\ket{-}_{1,2}$ subjected to $1/f$ noise arising from $N_f$ fluctuators with a distribution of frequency ratios ranging from $10^{-4}$ to $10^4$.
We see that the oscillations begin to disappear as the value of $N_f$ increases, suggesting a transition from non-Markovian to Markovian behaviour as $N_f$ increases. The non-Markovian measure \(\mathcal{N}_{\mathrm{BLP}}^{*}\) is also consequently expected to go to zero as $N_f \gtrsim 10$. (c) The performance of the binomial codes with $N=3$ and $K=8$ under the same $1/f$ noise as a function of $\tau$. We see oscillations which decay as $N_f$ increases. The behaviour for initial times is shown in the inset, where the infidelity is lower for smaller values of $N_f$.
}
\label{fig:1/f}
\end{figure*}
We further find a bound on the value of the fidelity for a pure dephasing channel. We note the following well-known inequality relating the distinguishability of states inferred from measurement outcomes to the trace distance between them~\cite{Nielsen_Chuang_2010}. For a pair of state, $\hat{\rho},\hat{\sigma}$, the $l_1$ distance between the probability distributions, $\{p_m:=\Tr(\hat{E}_m\hat{\rho})\}$ and $\{q_m:=\Tr(\hat{E}_m\hat{\sigma})\}$, for any set of POVMs, $\{\hat{E}_m\}$, is bounded from above by the trace distance between the two states:
\begin{align}
    \frac{1}{2}\sum_{m}\abs{p_m-q_m} \leq D(\hat{\rho},\hat{\sigma}).
\end{align}
In the present case, $\hat{\rho}\equiv \mathscr{N}(\ketbra{+}{+})$, $\hat{\sigma}\equiv \mathscr{N}(\ketbra{-}{-})$. $c_0^{\tau}$ and $c_1^{\tau}$ are the probability distribution of outcomes corresponding to any set of POVMs (here phase measurements, $\{\hat{M}(\phi):\phi\in [0,2\pi)\}$) that is used to distinguish the dual codewords that are going through the noisy channel, $\mathscr{N}$. Therefore, we have
\begin{align}
    \int_0^{2\pi}\dd \phi_1\; \abs{c^{\tau}_0-c^{\tau}_1}(\phi_1) \leq 2D(\ket{+}_N,\ket{-}_N),
\end{align}
where we note that $D(\ket{+}_N,\ket{-}_N$ depends on $\tau$.
As we have observed that the trace distance of the dual codewords of any rotation-symmetric bosonic encoding exhibits an oscillation with sharp dips, the corresponding dips in the fidelity of the Knill EC are also fundamentally bounded from above. Explicitly,
\begin{align}
    \mathcal{F}(\tau) \leq \frac{2+[1+D(\ket{+}_N,\ket{-}_N)][1+D(\ket{+}_M,\ket{-}_M)]}{6},
\end{align}
where we once again note that $D(\ket{+}_N,\ket{-}_N$ depends on $\tau$.
In Fig.~\ref{fig:fidelity}(c), we compare this bound to the numerical results of the fidelity obtained from the circuit. 

\section{Evolution of trace distance and infidelity as a function of time for binomial codes under $1/f$ noise}
\label{app:1/f}

In this section, we present the results of the analysis of the effect of $1/f$ noise on the codewords of the binomial code, in addition to the case of coherent states shown in Fig.~\ref{fig:main_results2}(c). Specifically, we consider a range of frequency ratios contributing to $1/f$ noise, spanning from $r = 10^{-4}$ to $r = 10^4$, and examine the evolution of the trace distance for various values of $N_f$, the number of fluctuators. Using a pair of binomial codewords with $N=1$ and $K=2$, $\ket{+}_{1,2}$ and $\ket{-}_{1,2}$, we observe that the amplitude of trace distance oscillations decreases as $N_f$ increases. This behavior is similar to the case where the initial states are coherent states (see Fig.~\ref{fig:1/f}(a)). In particular, for $N_f = 10$, the oscillations are nearly suppressed, suggesting that the non-Markovianity measure approaches zero as $N_f \gtrsim 10$.
Evaluating $\mathcal{N}_{BLP}^{*}$ for the binomial case requires increased computational effort as the trace distance decays after a very long time, the qualitative similarity to the coherent state case supports the conclusion that the non-Markovianity measure vanishes in the large-$N_f$ limit. 
In Fig.\ref{fig:1/f} (c), we show the results of the performance of the binomial code for the Knill-EC circuit with $N=3$ and $K=8$ under $1/f$ noise for the count of fluctuators $N_f= 1,2$ and $10$. As expected the infidelity shows oscillations for small values of $N_f$ displaying non-Markovian features and becomes monotonic for $N_f=10$ as a function of $\tau$. 
\bibliography{main}

\end{document}